\documentclass[preprint,onecolumn,superscriptaddress,amsmath,amssymb,aps,floatfix,nofootinbib,longbibliography]{revtex4-1}

\usepackage[pdftex]{graphicx}
\usepackage{bm}
\usepackage{color}
\usepackage{times}
\usepackage{amsmath}
\usepackage{amssymb}
\usepackage{mathtools}
\usepackage{gensymb} 
\newcommand{\defeq}{\vcentcolon=}
\usepackage{tablefootnote}
\usepackage{textcomp}
\usepackage{dsfont}

\usepackage[utf8]{inputenc}
\usepackage{natbib}
\usepackage{wrapfig}
\usepackage{lipsum}
\usepackage{xcolor}
\usepackage[normalem]{ulem}
\usepackage{amsfonts}
\usepackage{tikz, todonotes}
\usetikzlibrary{shapes,shadows,arrows}

\begin{document}

\title{A homogenised model for dispersive transport  and sorption in a heterogeneous porous medium}

\author{Lucy C. Auton}
\affiliation{Centre de Recerca Matem\`{a}tica, Campus de Bellaterra, Edifici C, 08193 Bellaterra, Barcelona, Spain}

\author{Mohit P. Dalwadi}
\affiliation{Mathematical Institute, University of Oxford, Oxford, OX2 6GG, UK}
\affiliation{Department of Mathematics, University College London, London WC1H 0AY, UK}

\author{Ian M. Griffiths}
\email{ian.griffiths@maths.ox.ac.uk}
\affiliation{Mathematical Institute, University of Oxford, Oxford, OX2 6GG, UK}

\date{\today}



\begin{abstract}
When a fluid carrying a passive solute flows quickly through porous media, three key macroscale transport mechanisms occur. These mechanisms are diffusion, advection and dispersion, all of which depend  on the microstructure of the porous medium; however, this dependence remains poorly understood. For idealised microstructures,  one can use the mathematical framework of homogenisation  to examine this dependence, but strongly heterogeneous materials are more challenging. Here, we consider a two-dimensional microstructure comprising an array of obstacles of smooth but arbitrary shape, the size and spacing of which can vary along the length of the porous medium. We use homogenisation via the method of multiple scales to systematically upscale a microscale problem involving non-periodic cells of varying area to obtain effective continuum equations for macroscale transport and sorption. The equations are characterised by the local porosity,  an effective local adsorption rate and an effective local anisotropic solute diffusivity. All of these macroscale properties depend nontrivially on the two degrees of microstructural geometric freedom in our problem; obstacle size and obstacle spacing. Further, the coefficient of effective diffusivity comprises the molecular diffusivity, the suppressive effect of the presence of obstacles and the enhancing effect of dispersion. To illustrate the mathematical model, we focus on a simple example geometry comprising circular  obstacles on a hexagonal lattice, for which we numerically determine the macroscale permeability and  effective diffusivity. We find a power law for the dispersive component of solute transport, consistent with classical Taylor dispersion.
\end{abstract}

\maketitle

\section{Introduction}
\label{s:intro}

Solute transport through porous materials is a fundamental process in many applications within biology, hydrogeology and environmental challenges such as contaminant transport and filtration (\textit{e.g.,} \cite{davit2013hydrodynamic, domenico1998physical,fritton2009fluid, mariani2010high, printsypar2019influence, spychala2015bacteria}). The majority of naturally occurring porous materials are intrinsically heterogeneous and/or anisotropic at the pore-scale and the macroscopic flow and transport are known to depend critically on the pore structure,  localised fluid--solid interactions and the connectivity of the fluid region \cite{beckwith2003anisotropy, wang2020effect}. For example, Rosti \textit{et al.} \cite{rosti2020breakdown} found that microstructural changes resulting from deformation of the solid phase of the porous material can cause a breakdown of Darcy's law. 
Despite the crucial role that microstructure plays in macroscale flow and solute transport, the significance of microscale geometry on these macroscale properties is often overlooked with most  models that systematically link microscale structure with macroscale transport  relying on a periodic  microscale structure.  Here, we consider the impact of microscale heterogeneity on macroscale  dispersion. We previously formally  derived a homogenised model to study the impact of slowly varying pore structure on macroscopic flow, transport and sorption within a porous medium \cite{auton2021homogenised}. 
Specifically, in Auton \textit{et al}. \cite{auton2021homogenised} we considered solute transport through a heterogeneous, two-dimensional porous material comprising an array of solid obstacles, where we allowed for slow but arbitrary longitudinal variations in the size and spacing of obstacles. The key difference between that work and the work here is that in Auton \textit{et al.} \cite{auton2021homogenised} diffusion dominates on the microscale so that no dispersive effects arise over the macroscale. In this manuscript, we are specifically interested in understanding emergent dispersive effects, and so advection becomes important on the microscale. Here, we develop a homogenised model for dispersive transport through the same heterogeneous porous material. 

The P\'{e}clet number is defined as the ratio of advective transport to diffusive transport and thus its size determines which transport mechanisms dominate. There are two  P\'{e}clet numbers of particular interest: the local P\'{e}clet number $\mathrm{Pe}_l$, based on the pore size, and the global P\'{e}clet number $\mathrm{Pe}_g$, based on the size  of the solute pulse.
Dispersion arises at the macroscale in the limit where advection and diffusion balance at the microscale ($\mathrm{Pe}_l=\mathcal{O}(1)$),  so that, at the macroscale, advection dominates ($\mathrm{Pe}_g\gg1$). This limit is crucial for many environmental and industrial applications for example in the formation and functioning of wetland systems, bacteria or virus transport in ground water, water injection into oil reserves,  and industrial filtration (\textit{e.g.,} \cite{bedrikovetsky2006correction,huang2008advection}). 

The first investigations into dispersion were conducted in the 1950s by Taylor~\cite{taylor1953dispersion} and Aris~\cite{aris1956dispersion}. Taylor and Aris investigated solute transport through a tube (Poiseuille flow) and derived  an asymptotic equation for the average cross-sectional concentration in the tube, finding that the dispersive component of transport is asymptotically  proportional to the square of the P\'{e}clet number.  Experimentally, a range of power laws have been found relating P\'{e}clet number and dispersion; Dronfield \& Silliman \cite{dronfield1993velocity} find that for smooth parallel plates this power law is obtained but for rougher surfaces the exponent decreases, as the boundary effects become more dominant and different dispersion mechanisms dominate. 

Macroscale dispersion in the case of homogeneous porous materials (\textit{i.e.,} porous materials which comprise a periodic, repeating microstructure)   has previously been obtained using homogenisation via the method of multiple scales (MMS). Salles \textit{et al.} \cite{salles1993taylor} compare different methods for theoretically deriving the dispersion tensor: the method of moments, homogenisation via the MMS, using both multiple spatial and temporal scales, and a purely numerical approach based on random walks. An alternative homogenisation approach employs a drift transformation conducted simultaneously with the homogenisation~\cite{allaire2010two, allaire2007homogenization}. Drift transformations have been used  within formal asymptotic investigations into Taylor dispersion \cite{griffiths2013control}. 
Davit \textit{et al.} \cite{davit2013homogenization} give a comparison of the different  upscaling methodologies:  homogenisation via the method of multiple scale and volume averaging. 

We expect to see a rich transport behaviour  when the dispersive component dominates, which corresponds to high $\text{Pe}_l$. Liu \textit{et al.} \cite{liu2024non} note non-monotonicity of dispersion in a compressed spherical packing of elastic spheres, with respect to varying particle P\'{e}clet number. Further, Liu \textit{et al.} \cite{liu2024non} explain that as the P\'{e}clet number varies  the macroscale dispersion is dominated by different dispersion and diffusion mechanisms: molecular diffusion, which dominates at small $\text{Pe}_l$, hold-up dispersion (areas of no/slow flow),  shear (Taylor) dispersion caused by non-uniform velocity profiles within pores or throats, and mechanical dispersion that results from the repeated separation and merging of flow passages at the junctions of the pore space. This competition amongst the different mechanisms leads to a rich and varied behaviour. 

Here, we investigate the effect of a slowly varying microstructure on dispersive transport of a solute pulse.
In particular, we formally derive a homogenised model for dispersive transport through a heterogeneous porous material comprising an array of arbitrarily shaped obstacles with two degrees of microstructural freedom. Firstly, we allow for the obstacle size to isotropically varying along the length of the porous material and, additionally, we allow for the spacing between obstacles vary along the length of the porous material. 
We derive the transport results for a flow field under fairly general assumptions (incompressible flow subject to no-slip and no-penetration conditions on the obstacle surfaces), and subsequently we explicitly calculate the flow field and its effect on dispersion for the example of Stokes flow.
To focus on the effects of dispersion  we consider transient advection and diffusion with removal via adsorption on the solid surfaces such that advection dominates on the macroscale (\S\ref{s:rad}). 
The variation in the spacing between obstacles means standard homogenisation techniques  cannot be applied to this problem and thus we present a novel, modified, homogenisation to accommodate this microscale heterogeneity. 
Following the homogenisation methodology laid out in Salles \textit{et al}., \cite{salles1993taylor}
we introduce a second, fast timescale that balances with the dominant advective term. 
We exploit the local periodicity of the pore geometry to formally homogenise the pore-scale problem via the MMS \cite{dalwadi2016multiscale, dalwadi2015understanding, bruna2015diffusion}.  Subsequently, we perform a drift transformation to separate the leading-order effect of the solute pulse advecting with the flow from the spreading around this moving frame of reference (\S\ref{s:homog}). 
The homogenisation method provides macroscale equations that are uniformly valid over the entire porous medium. For a particular cell geometry the effective diffusivity tensor, which comprises components due to molecular diffusion, a reduction in spreading due to the presence of obstacles, and dispersion, must be determined numerically. To demonstrate the general approach, we choose a particular example geometry comprising a hexagonal array of circular obstacles, and determine the effective diffusivity tensor for individual cells with a wide range of obstacle size and spacing (\S\ref{exs}). For this example geometry, the dispersion is shown to depend on the square of the P\'{e}clet number~\cite{aris1956dispersion, davit2013homogenization, taylor1953dispersion}. Finally, we discuss the merits and limitations of the model~(\S\ref{s:conclusion}).

\section{Model Problem}
\label{s:rad}

\begin{figure}
    \centering
    \includegraphics[width=\textwidth]{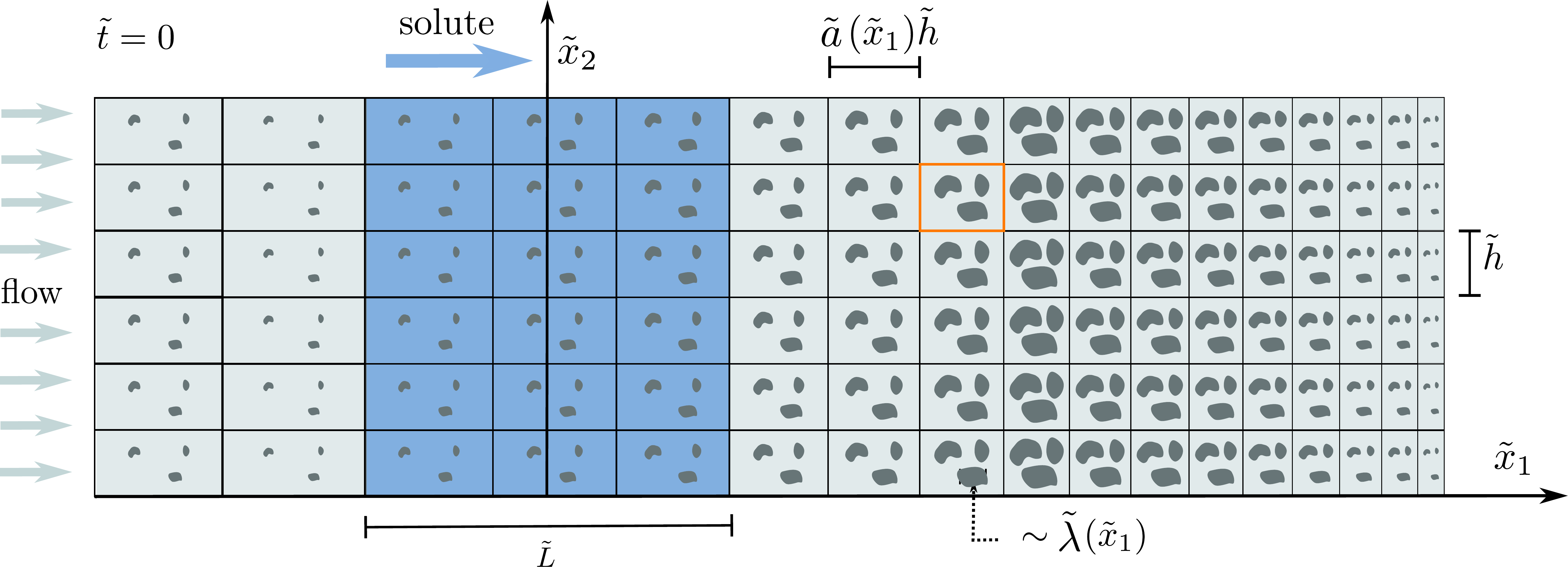}
    \caption{ \label{fig:schem_dim_1}
    We consider the flow of fluid carrying solute through a heterogeneous porous filter in two dimensions.
The porous medium is formed of an array of obstacles whose size depends only on a scale factor $\tilde{\lambda}(\tilde{x}_1)$, located within each rectangular cell
    of constant transverse height $\tilde{h}$, and longitudinal width $\tilde{a}(\tilde{x}_1)\tilde{h}$. The porous medium is thus uniform in the transverse ($\tilde{x}_2$) direction but heterogeneous in the longitudinal ($\tilde{x}_1$) direction. We assume that the spacing between obstacles is small relative to the initial length of solute pulse, $\tilde{L}$, contained within the porous medium  --- that is,  $\epsilon\defeq\tilde{h}/\tilde{L} \ll 1$. We isolate one cell in orange; a dimensionless version of this cell is shown in detail in Figure \ref{Fig_schem_2}. 
}
\end{figure}

We consider a porous material with the same  microstructural freedom as the porous media developed in Auton \textit{et al}. \cite{auton2021homogenised}, but now with  flows that are suitably fast that dispersive transport is non-negligible at the macroscale. In particular, we consider the steady flow of fluid carrying a passive solute through a rigid porous medium in two dimensions. The solute advects, diffuses, disperses and is removed via adsorption to the solid structure.  The spatial coordinate is
$\tilde{\bm{x}}\defeq\tilde{x}_1\bm{e}_1+\tilde{x}_2\bm{e}_2$, with $\tilde{x}_1$ and $\tilde{x}_2$ the dimensional longitudinal and transverse coordinates, respectively, and $\bm{e}_1$ and $\bm{e}_2$ the longitudinal and transverse unit vectors, respectively. The porous material is of infinite extent, in both the $\tilde{x}_1$ and $\tilde{x}_2$  directions.  We consider a solute pulse of initial length $\tilde{L}$ being advected along the length of the porous material. At time $\tilde{t}=0$, we fix the $\tilde{x}_1$-origin, defined by $\tilde{x}_1=0$, to the centre of the solute pulse.

The entire domain of the porous medium, denoted $\tilde{\Omega}$, comprises both the fluid and the solid structure of the domain. 
The latter constitutes an array of solid obstacles, as discussed in more detail below.
We assume that the solute particles are negligibly small relative to the solid obstacles, and we measure the local density of solute (amount of solute per volume of fluid in kg$/\mathrm{m}^3$) via the concentration field $\tilde{c}(\tilde{\bm{x}},\tilde{t})$. 
This concentration field is defined within the fluid phase of the porous medium, denoted $\tilde{\Omega}_f$.

Note that we do not track solute once it has adsorbed to the solid surface, and we neglect any impact of this adsorption on the size of the obstacles. The latter point is justified by our assumption that the solute particles are negligible in size relative to the obstacles, and also because we are interested in macroscopic diffusive and advective timescales, which are typically far shorter than those of solute accumulation and blocking.

The porous medium can be partitioned into an array of rectangular cells of fixed height~$\tilde{h}$ and varying width $\tilde{a}(\tilde{x}_1)$, where $\tilde{a}$ is the aspect ratio of a given cell and where $\tilde{x}_1$ is taken be the centre of each cell.
Each cell contains fixed and rigid obstacles of smooth but arbitrary shape. The shape of each obstacle is fixed and each obstacle can only grow or shrink isotropically about their respective centre of mass according to a scale factor $\tilde{\lambda}(\tilde{x}_1)$.  The solid domain is the union of these obstacles, and is denoted $\tilde{\Omega}_s\defeq \tilde{\Omega}\setminus\tilde{\Omega}_f$. 
This construction leads to a porous medium whose properties vary in the longitudinal direction but not in the transverse direction (see Figure~\ref{fig:schem_dim_1}). We further assume that the length of the initial solute pulse is much greater than the height of each cell comprising the porous medium, which requires $\epsilon \ll 1$ where we define   $\epsilon\defeq \tilde{h}/\tilde{L}$.     This formulation allows the porous medium to have $\mathcal{O}(1)$ variations in microstructure over an $\mathcal{O}(1)$ variation in $\tilde{\bm{x}}$.

We model solute transport and adsorption via the standard advection--diffusion equation with a linear, partially adsorbing condition at the fluid--solid interface:
\begin{subequations}\label{ad_diff_dim}
    \begin{align}
    \label{c_hat_dim}
        \frac{\partial {\tilde{c}}}{\partial \tilde{t}} &= {\tilde{\boldsymbol{\nabla}}}\cdot\left(\tilde{{D}}\tilde{{\boldsymbol{\nabla}}}{\tilde{c}}- \tilde{\bm{v}}\tilde{c}\right), \ \quad \tilde{\bm{x}}\in\tilde{\Omega}_f,\\
    \label{Robin_dim}
        -\tilde{\gamma}\hat{c} &= \tilde{\bm{n}}_s\cdot\left(\tilde{D}\tilde{{\boldsymbol{\nabla}}}\tilde{c}-{\tilde{\bm{v}}}\tilde{c}\right), \quad \tilde{\bm{x}}\in \partial\tilde{\Omega}_s, 
    \end{align}
\end{subequations}
where $\tilde{{D}}$ is the coefficient of molecular diffusion,  $\tilde{\bm{v}}$  is the given fluid velocity (\textit{e.g.,} see \S\ref{sec_flow_assumps}), 
 $\tilde{\bm{n}}_s$ is the outward-facing unit normal to $\partial\tilde{\Omega}_s$, and $\tilde{\gamma} \geq 0$ 
is the constant adsorption coefficient. Further, we note that $\tilde{\gamma} = 0$ corresponds to no adsorption and $\tilde{\gamma}\to\infty$ corresponds to instantaneous adsorption, where the latter is equivalent to imposing $\tilde{c} = 0$ on $\partial\tilde{\Omega}_s$.

To deal with the boundaries during the upscaling, it is helpful to define a function $\tilde{f}_s(\tilde{\bm{x}})$ such that on the fluid--solid interface $\partial \tilde{\Omega}_s$
\begin{equation}
 \tilde{f}_s(\tilde{\bm{x}}) = 0. 
\end{equation}
We also define $\tilde{f}_s(\tilde{\bm{x}})>0$ inside the solid phase. Then, 
\begin{equation}
\label{nhat_dim}
    \tilde{\bm{n}}_s(\tilde{\bm{x}}) \defeq \frac{\tilde{\bm{\nabla}}\tilde{f}_s}{\left|\tilde{\bm{\nabla}}\tilde{f}_s\right|},
\end{equation}
is the outward-facing normal to the fluid domain.

\subsection{Dimensionless Equations}
We make Equations (\ref{ad_diff_dim}) dimensionless via the scalings
\begin{equation}\label{eq:non-dimensionalisation}
    \tilde{\bm{x}} = \tilde{L}\hat{\bm{x}}, \quad \tilde{\bm{v}} = \tilde{{V}}\hat{\bm{v}}, 
    \quad \tilde{c} = \tilde{{C}} \hat{c}, \quad\text{and}\quad \tilde{t} = \left(\frac{\tilde{L}^2}{\tilde{{D}}}\right)\hat{t},
\end{equation}
where $\tilde{{{V}}}$ and $\tilde{{{C}}}$ are the average inlet velocity and the average inlet concentration, respectively; $\bm{\hat{x}}$ and $\hat{t}$ denote the dimensionless spatial and temporal coordinates, respectively; and $\hat{\bm{v}}= \hat{\bm{v}}(\bm{\hat{x}})$ and $\hat{c}= \hat{c}(\bm{\hat{x}},\hat{t})$ denote the dimensionless velocity  and concentrations fields, respectively.

Employing the scalings in Equation~(\ref{eq:non-dimensionalisation}),  the transport problem (Eqs.~\ref{ad_diff_dim}) becomes 
\begin{subequations}\label{ad_diff}
    \begin{align}
    \label{c_hat}
        \frac{\partial {\hat{c}}}{\partial \hat{t}} &= {\hat{\boldsymbol{\nabla}}}\cdot\left(\hat{{\boldsymbol{\nabla}}}{\hat{c}}-\mathrm{Pe}_l \frac{{\hat{\bm{v}}}\hat{c}}{\epsilon}\right), \ \quad \hat{\bm{x}}\in\hat{\Omega}_f,\\
    \label{Robin}
        -\epsilon\gamma\hat{c} &= \hat{\bm{n}}_s\cdot\left(\hat{{\boldsymbol{\nabla}}}\hat{c}-\mathrm{Pe}_l\frac{{\hat{\bm{v}}}\hat{c}}{\epsilon}\right), \quad \hat{\bm{x}}\in \partial\hat{\Omega}_s. 
    \end{align}
\end{subequations}
Here, ${\hat{\boldsymbol{\nabla}}}$ is the gradient operator with respect to the spatial coordinate $\hat{\bm{x}}$, $\hat{\bm{n}}_s(\hat{\bm{x}})$ is the outward-facing normal to $\hat{\Omega}_f$,  
 the dimensionless adsorption rate $\gamma\defeq{\tilde{\gamma} \tilde{L}}/{(\epsilon \tilde{\mathcal{D}})}= \mathcal{O}(1)$ (to obtain a distinguished limit), measures the rate of adsorption relative to that of diffusive transport, and the
local P\'{e}clet number 
\begin{equation}
\label{Pes}
\mathrm{Pe}_l\defeq{\tilde{h}\tilde{\mathcal{V}}}/\tilde{\mathcal{D}}\equiv\epsilon\mathrm{Pe}_g,   
\end{equation}   
where the local P\'{e}clet number
measures the rate of advective transport relative to that of  diffusive transport across each cell while the global P\'{e}clet number $\mathrm{Pe}_g\defeq {\tilde{L}\tilde{\mathcal{V}}}/\tilde{\mathcal{D}}$ measures the rate of advective transport relative to that of  diffusive transport across an $\mathcal{O}(1)$  section of the filter. 
Note that dispersive effects appear in the leading-order macroscale transport equation when the global P\'{e}clet number is of  $\mathcal{O}(1/\epsilon)$, so that $\mathrm{Pe}_l=\mathcal{O}(1)$. 

Finally, the dimensionless fluid--solid interface becomes $\hat{f}_s(\hat{\bm{x}}) = 0$ and Equation (\ref{nhat_dim}) becomes 
\begin{equation}
\label{nhat}
    \hat{\bm{n}}_s(\hat{\bm{x}}) \defeq \frac{\hat{\bm{\nabla}}\hat{f}_s}{\left|\hat{\bm{\nabla}}\hat{f}_s\right|}. 
\end{equation}

\subsection{Method of multiple scales}
Following the method of multiple scales (MMS), we isolate and solve the  solute transport problem  in an individual cell, which is characterised by its aspect ratio 
\begin{equation}
    \hat{a}(\hat{x}_1)=\tilde{a}(\tilde{x}_1), 
\end{equation}
and obstacle scale factor 
\begin{equation}
    \hat{\lambda}(\hat{x}_1)= \tilde{\lambda}(\tilde{x}_1). 
\end{equation}
We then construct a model for macroscopic flow and transport through the entire porous medium from the solution to these individual cell problems via local averaging. 
The result is a system of equations that are uniformly valid for all $\hat{\bm{x}}\in\hat{\Omega}$.
\subsubsection{Spatial transform}
\begin{figure}
    \centering
    \includegraphics[width=\textwidth]{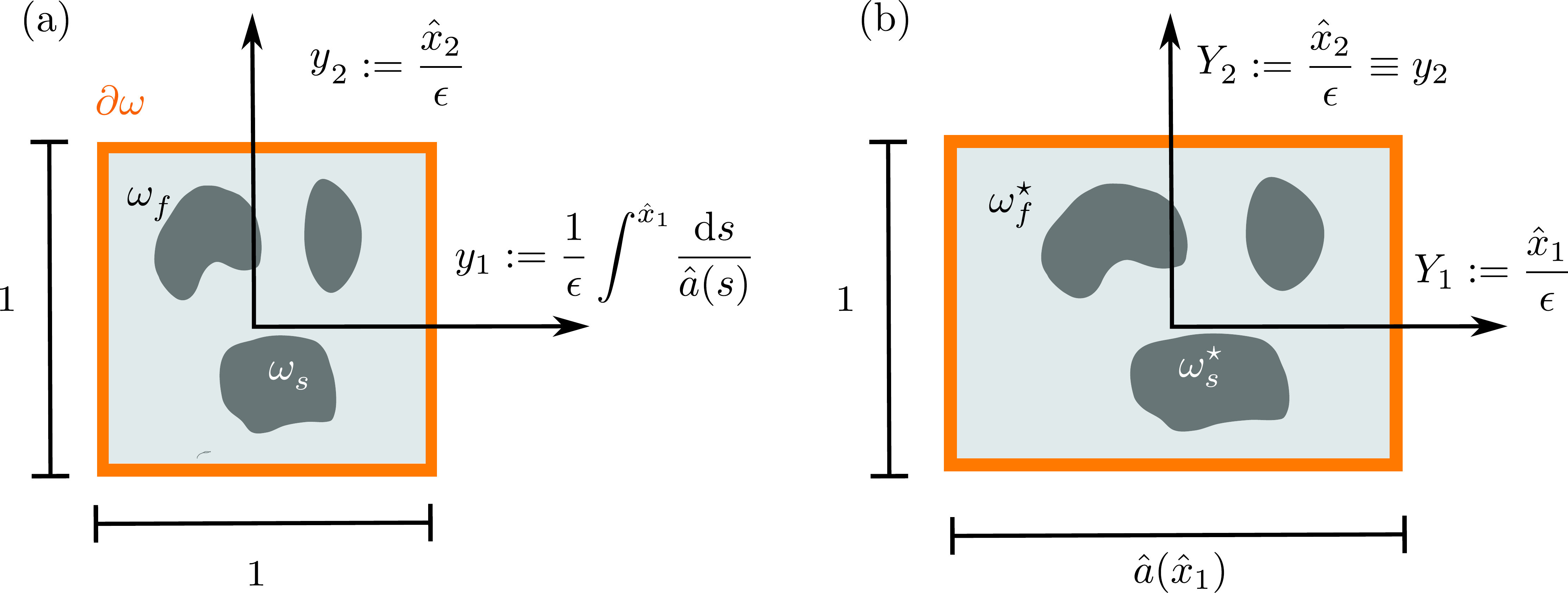}
    \caption{
    An arbitrary cell within the porous medium (orange rectangle in Figure~\ref{fig:schem_dim_1}) represented in (a)~transformed microscale coordinates and (b)~physical microscale coordinates. We map the physical microscale coordinates $Y_1$ and $Y_2$ to transformed microscale coordinates $y_1$ and $y_2$ according to Equations~\eqref{y_map} and \eqref{Y_map} to scale the slow variation in cell width $a$ out of the cell problem, such that each physical rectangular cell is transformed into a square. Note that $\partial\omega \defeq \partial \omega_{||}\cup\partial\omega_{=}$, where $\partial \omega_{||}$ and $\partial\omega_{=}$ denote the vertical and horizontal cell boundaries, respectively, and that  the domains and boundaries in the (physical) rectangular $\bm{Y}$-cell will be denoted as in the square $\bm{y}$-cell, but with the addition of a superscript~$\star$. \label{Fig_schem_2}
     }
\end{figure}

A consequence of the obstacle size and spacing varying in the $\hat{x}_1$ direction is that  the period of the fast scale varies over the slow scale, thus, we cannot use standard homogenisation techniques here. Instead, we follow the approach from Auton \textit{et al}. \cite{auton2021homogenised}, based on previous methodology developed in Chapman and McBurnie \cite{chapman2011unified}, and Richardson and Chapman \cite{richardson2011derivation} and  
define both a transformed microscale coordinate $\bm{y}=(y_1,y_2)$ given by
\begin{equation}\label{y_map}
    {y_1}\defeq \frac{1}{\epsilon} \int^{\hat{x}_1} \! \frac{\mathrm{d}s}{\hat{a}(s)} \quad \text{and} \quad y_2 \defeq \frac{\hat{x}_2}{\epsilon},
\end{equation}
for which  each cell is of unit volume,
and a physical microscale coordinate $\bm{Y}=(Y_1,Y_2)$ given by
\begin{equation}
\label{Y_map}
    \bm{Y} = \frac{\hat{\bm{x}}}{\epsilon}
\end{equation}
for which the total cell volume is $\hat{a}(x_1)$ and where the shape of the obstacles remains unchanged. The benefit of introducing $\boldsymbol{y}$ is that the microscale cell size does not change over the macroscale in this coordinate. Therefore, we may safely perform a homogenisation in $\boldsymbol{y}$. We will switch to $\bm{Y}$ to calculate specific integrals that arise during the homogenisation procedure as this domain is more straightforward to work with numerically. 

Following the MMS, we define a macroscale spatial coordinate  $\bm{x}\defeq\hat{\bm{x}}$ and we take $\bm{x}$ and $\bm{y}$ to be independent spatial parameters. Thus spatial derivatives become 
\begin{subequations}
\label{spatial_map}
\begin{align}
\frac{\partial}{\partial \hat{x}_i} = \frac{\partial}{\partial x_i} + \frac{\sigma_{ij}}{\epsilon} \frac{\partial}{\partial y_j},
\end{align}
for $i,j = 1, 2$, and where $\sigma_{ij}\defeq\left(\bm{\sigma}\right)_{ij}$ and 
\begin{equation}
\label{sigma}
    \bm{\sigma}=\begin{pmatrix}
 \displaystyle\frac{1}{\hat{a}(x_1)} & 0\\
 0 & 1
\end{pmatrix}.
\end{equation}
Alternatively, in vector form, the spatial derivatives become
\begin{align}
\hat{\bm{\nabla}} \defeq \bm{\nabla}_x+\frac{1}{\epsilon}\bm{\nabla}_y^a
\end{align}
where $\boldsymbol{\nabla}_x$ is the gradient operator with respect to the coordinate $\bm{x}$ and where
\begin{equation}
    \bm{\nabla}_y^a\defeq \left(\frac{1}{\hat{a}}\frac{\partial}{\partial y_1},\frac{\partial}{\partial y_2}\right)^\intercal,
\end{equation}
 is the gradient operator associated with the $\bm{y}$-coordinate transform. 
 \end{subequations}

 \subsubsection{Temporal scales}
 For a porous material with a fixed microstructure,  classically one would conduct the homogenisation in a frame that advects  with the flow (\textit{cf}. \cite{allaire2010two, allaire2007homogenization}). 
 However due to the slowly varying microscale geometry in this problem, it is preferable to introduce  two timescales: a fast advective timescale, $\tau = \hat{t}/\epsilon$, which tells us which frame to move into, and a slower diffusive/dispersive timescale,  $ t =\hat{t} $, which allows us to quantify and characterise the dispersive effects in which we are interested. This approach is motivated by Salles \textit{et al}. \cite{salles1993taylor}. 
 Thus the time derivatives become 
\begin{equation}
\label{Eq_2ts}
    \frac{\partial}{\partial\hat{t}} =  \frac{\partial}{\partial{t}}+ \frac{1}{\epsilon}\frac{\partial}{\partial{\tau}}.
\end{equation}

 We therefore rewrite all functions of $\hat{\bm{x}}$ and $\hat{t}$  as functions of $\bm{x}$ and  $\bm{y}$, and $t$ and $\tau$, respectively:
$\hat{\bm{v}}(\hat{\bm{x}})\defeq\bm{v}(\bm{x},\bm{y})$,  
 and $\hat{c}(\hat{\bm{x}},\hat{t}) \defeq c(\bm{x},\bm{y}, t, \tau)$. 
 If we are referring to functions of $\bm{Y}$ in lieu of $\bm{y}$, we adorn each function with a superscript $\star$.   
 We define the domains for the microscale cells as in Figure~\ref{Fig_schem_2}, and  the    solid-fluid boundary as $\partial\omega_s$ in the square $\bm{y}$-cell  and as $\partial\omega_s^\star$ in the (physical) rectangular $\bm{Y}$-cell.

\subsection{\textbf{Flow assumptions}}  
\label{sec_flow_assumps}

Our focus in this work is deriving the dispersive contaminant transport through heterogeneous porous media. Since our derivation allows for a general flow field under fairly minimal assumptions, we lean into this generality and derive the transport results for a flow field under the following minimal assumptions:
\begin{enumerate}
\item The flow is steady and $\hat{\bm{v}}$ is bounded within each cell. 
\item The fluid is incompressible --- that is: 
\begin{subequations}
\label{flow_assumps}
\begin{equation}
\label{incomp_assump}
\hat{\bm{\nabla}}\cdot\hat{\bm{v}}=0, \qquad \hat{\bm{x}}\in \hat{\Omega}_f \end{equation}
\item On the boundary of the solid obstacles, we have a no-slip and no-penetration condition:
 \begin{equation}
 \label{noslipnopen_assump}
\hat{\bm{v}}=0, \qquad \hat{x}\in\partial\hat{\Omega}_s.
\end{equation}
\end{subequations}
\item The periodic microscale results in the flow being locally periodic over the microscale (\textit{cf.}  \cite{allaire2010two, allaire2007homogenization, auton2021homogenised}).
\end{enumerate}
In essence the flow field can come from any type of incompressible flow \cite{allaire2010two, allaire2007homogenization}. For Stokes flow,  the   homogenisation of the flow problem with this general microscale geometry  is considered in Auton \textit{et al}. \cite{auton2021homogenised}. We will also explicitly calculate the flow field and its effect on dispersion later for the example of Stokes flow, in \S\ref{exs}.

\subsection{Averaging} 

 For a given quantity $Z(\bm{x},\bm{y},t, \tau) = Z^\star(\bm{x},\bm{Y},t, \tau)$, there are two different averages of interest: the intrinsic (fluid) average
\begin{multline}
\label{bra_ket_def}
    \langle Z\rangle(\bm{x},t, \tau) \defeq  \frac{1}{|\omega_f(x_1)|} \! \int_{\omega_f(x_1)} Z(\bm{x},\bm{y},t, \tau) \, \mathrm{d}S_y \equiv \\ \frac{1}{|\omega_f^{\star} (x_1)|}\! \int_{\omega_f^\star (x_1)} Z^\star(\bm{x},\bm{Y},t, \tau) \, \mathrm{d}S_Y = \langle Z^\star\rangle(\bm{x},t, \tau),
\end{multline}
where the total fluid area in the transformed cell $|\omega_f|$ (or physical cell $|\omega_f^{\star}|$) is a function of $\hat{a}(x_1)$ and $\lambda(x_1)$;
and the volumetric average
\begin{equation}
\label{volume_int_y2Y}
    \frac{1}{|\omega(x_1)|}\int_{\omega(x_1)} \! Z(\bm{x},\bm{y},t, \tau)\, \mathrm{d} S_y \equiv \frac{1}{|\omega^{\star}(x_1)|}\int_{\omega^{\star}(x_1)} \!  Z^\star(\bm{x},\bm{Y},t,, \tau) \, \mathrm{d}S_Y, 
\end{equation}
where $|\omega|=1$ and $|\omega^{\star}|=\hat{a}$.  
Here, $\mathrm{d}S_y\defeq\mathrm{d}y_1\mathrm{d}y_2$ is an area element of the transformed microscale fluid region, $\mathrm{d}S_Y\defeq\mathrm{d}Y_1\mathrm{d}Y_2$ is an area element of the physical microscale fluid region and the porosity $\hat{\phi}$ is 
\begin{equation}
\label{phi}
\hat{\phi}(x_1) = \frac{|\omega_f(x_1)|}{|\omega(x_1)|}\equiv  |\omega_f(x_1)| \left(= \frac{|\omega_f^{\star}(x_1)|}{|\omega^{\star}(x_1)|} \right),
\end{equation}
since $|\omega(x_1)|\equiv 1$ by construction. 
Thus, the intrinsic average $\langle c\rangle$ is the amount of solute per unit fluid area 
within the porous medium, while $\hat{\phi}\langle c\rangle$, the volumetric average of the concentration, is the amount of solute per unit total area. We will use the intrinsic average (Eq.~\ref{bra_ket_def}) in the work that follows.

\section {Homogenisation}
\label{s:homog}
\subsection{Incompressible flow}
Here, we use incompressibility of flow and its boundary condition to determine some micro- and macroscale flow expressions.  In particular,  we expand the fluid velocity field asymptotically as follows: 
\begin{equation}
\label{eq:v_expansion}
\bm{v}(\bm{x},\bm{y}) = \bm{v}^{(0)}(\bm{x},\bm{y})+\epsilon\bm{v}^{(1)}(\bm{x},\bm{y})+\epsilon^2\bm{v}^{(2)}(\bm{x},\bm{y})+\cdots \quad \text{as}\quad \epsilon\to0. 
\end{equation}
Using the assumptions detailed in \S\ref{sec_flow_assumps}, we 
apply the multiple spatial scales  ansatz (Eq.~\ref{spatial_map}) and the expansion (Eq.~\ref{eq:v_expansion}) to Equations (\ref{flow_assumps}) and compare orders of $\epsilon$ yielding a cascade of equations  in 
orders of $\epsilon$: 
\begin{subequations}
\label{flow_core}
\begin{align}
\label{O0incomp}
\bm{\nabla}_y^a\cdot\bm{v}^{(0)} = 0, \qquad &\bm{y}\in\omega_f, \\ 
\label{Oiincomp}
    \bm{\nabla}_x\cdot\bm{v}^{(r)} =-\bm{\nabla}_y^a\cdot\bm{v}^{(r+1)},\qquad & \bm{y}\in\omega_f, \quad\text{for }\ r \in \mathbb{N}_{\geq0},  \\
    \label{vi}
\bm{v}^{(r)} = \bm{0}, \qquad &\bm{y}\in\partial\omega_s \quad\text{for }\ r \in \mathbb{N}_{\geq0},
\end{align}
where  Equation (\ref{O0incomp}) tells us that the leading-order microscale velocity is incompressible on the microscale and Equation (\ref{vi}) yields   a no-slip and no-penetration condition on the solid obstacles at all orders. The Dirichlet (no slip) boundary condition means that the full transformation machinery is not required for this part of the analysis.
Further, using flow assumption 4 (periodicity of $\bm{v}$ over the microscale; see \S\ref{sec_flow_assumps}) and expansion (Eq.~\ref{eq:v_expansion}), we determine
\begin{equation}
\label{flow_period}
{v}_i^{(r)}, \quad \text{periodic on} \quad \bm{y}\in\partial\omega_{=}
(x_1)\ \ \text{and}\ \  \partial\omega_{||}(x_1),  \quad\text{for }\ r \in \mathbb{N}_{\geq0} \ \ \text{ and }\ \ i\in\{1,2\}.
\end{equation}
\end{subequations}

We consider the intrinsic average (Eq.~\ref{bra_ket_def}) of Equation (\ref{Oiincomp}); 
\begin{multline}
\label{incomp_intrinsic_ave}
\frac{1}{|\omega_f(x_1)|} \ \int_{\omega_f(x_1)} \bm{\nabla}_x\cdot\bm{v}^{(r)}\ \mathrm{d}S_y = \\-\frac{1}{|\omega_f(x_1)|} \ \int_{\omega_f(x_1)}\bm{\nabla}_y^a\cdot\bm{v}^{(r+1)}\mathrm{d}S_y , \qquad  \bm{y}\in\omega_f, \quad\text{for }\ r \in \mathbb{N}_{\geq0},
\end{multline}
To manipulate the first term on the right-hand side of Equation (\ref{incomp_intrinsic_ave}) we use the transport theorem
\begin{subequations}
\begin{equation}\label{transport_final}
    \int_{\omega_f} \! \bm{\nabla}_x\cdot\bm{z}\, \mathrm{d}S_y =     \bm{\nabla}_x\cdot\int_{\omega_f} \! \bm{z}\, \mathrm{d}S_y + \int_{\partial\omega_s} \! \bm{N}\cdot\bm{z}\, \mathrm{d}s_y,
\end{equation}
 where the macroscale perturbation to the normal $\bm{N} = N_i \bm{e}_i$ is defined as
\begin{equation}\label{normal_N}
    \bm{N}\defeq\frac{\bm{\nabla}_xf_s}{|\bm{\nabla}_y f_s|}. 
\end{equation}
\end{subequations}
The transport theorem (Eq.~\ref{transport_final}) is derived in Appendix A of Auton \textit{et al}. \cite{auton2021homogenised}.

Thus, we  apply the  transport theorem (Eq.~\ref{transport_final}) to the left-hand side of Equation (\ref{incomp_intrinsic_ave}) and use the no-slip and no-penetration conditions at the $i^\text{th}$ order (Eq.~\ref{vi}) so that Equation~(\ref{incomp_intrinsic_ave}) becomes 
\begin{multline}
\label{incomp_intrinsic_ave2} 
\frac{1}{|\omega_f(x_1)|} \ \bm{\nabla}_x\cdot \int_{\omega_f(x_1)} \bm{v}^{(r)}\ \mathrm{d}S_y = \\ -\frac{1}{|\omega_f(x_1)|} \ \int_{\omega_f(x_1)}\bm{\nabla}_y^a\cdot\bm{v}^{(r+1)}\mathrm{d}S_y , \qquad  \bm{y}\in\omega_f, \quad\text{for }\ r \in \mathbb{N}_{\geq0},
\end{multline}
Applying the the divergence theorem to  the right-hand side of Equation (\ref{incomp_intrinsic_ave2}) combined with the solid boundary condition (Eq.~\ref{vi}), periodicity (Eq.~\ref{flow_period})  and the definition of the intrinsic average (Eq.~\ref{incomp_intrinsic_ave}) yields 
\begin{multline}
\label{macro_incomp} 
\frac{1}{|\omega_f(x_1)|} \ \bm{\nabla}_x\cdot \int_{\omega_f(x_1)} \bm{v}^{(r)}\ \mathrm{d}S_y \equiv \frac{1}{|\omega_f(x_1)|}\  \bm{\nabla}_x\cdot  \langle \hat{\phi}\bm{v}^{(r)}\rangle= 0  , \qquad  \bm{y}\in\omega_f, \quad\text{for }\ r \in \mathbb{N}_{\geq0}.
\end{multline}
Equation~(\ref{macro_incomp}) corresponds to incompressibility over the macroscale  at all asymptotic orders. Note that Equations (\ref{flow_core}) and~(\ref{macro_incomp}) are  not sufficient to define the flow field, however they do give us relations on both the micro- and macroscales, which we will use for the homogenisation of the solute transport problem. 

\subsection{Solute transport with dispersion}
We now homogenise the solute transport problem (Eqs.~\ref{ad_diff}). While the setup is similar to that of \S3.3 of Auton \textit{et al}. \cite{auton2021homogenised}, the key difference is the importance of dispersive effects. Among other technical differences, this also requires the introduction of a second temporal scale to allow for the systematic derivation of the macroscale solute-transport equation with dispersive effects. In particular, we  first determine a fast evolution equation for the solute due to advection;  then we consider higher-order equations in the concentration problem to determine a transport equation that balances all transport mechanisms. Finally, we combine these results to obtain an advection--diffusion equation which accounts for the emergent dispersion.

 \subsubsection{Treatment of the normal to the solid}
 Under the multiple scales framework, the unit normal to the solid interface is written as a function of both the macro- and microscales: $\hat{\bm{n}}_s(\hat{\bm{x}}) = \bm{n}_s(\bm{x},\bm{y})$, and similarly for the function $\hat{f}_s(\hat{\bm{x}}) = f_s(\bm{x},\bm{y})$, which vanishes on the solid interface. As in Auton \textit{et al}. \cite{auton2021homogenised}, to determine the correct form of  $\bm{n}_s(\bm{x},\bm{y})$ we must apply  the multiple-scales transformations (Eqs.~\ref{spatial_map}) consistently to Equation (\ref{nhat_dim}) which  yields
 \begin{subequations}
 \label{normals}
 \begin{align}
\label{n transform in geo} 
    \bm{n}_s =
    \frac{\displaystyle \left(\sigma_{ij} n^y_j + \epsilon N_i\right) \bm{e}_i}{\left[\displaystyle \sigma_{kl}\sigma_{km} n^y_ln^y_m \right]^{1/2} + \mathit{O}(\epsilon)},
\end{align}
where the geometric normal to the transformed solid obstacles $\bm{n}^y= n_i^y \bm{e}_i$ is given by
 \begin{align}
    \label{eq: n geo} 
 \bm{n}^y := \frac{\boldsymbol{\nabla}_y f_s}{\left|\boldsymbol{\nabla}_y f_s \right|} = \frac{\displaystyle \frac{\partial f_s}{\partial y_i}  \bm{e}_i}{\left[\displaystyle \frac{\partial f_s}{\partial y_j} \frac{\partial f_s}{\partial y_j} \right]^{1/2}}.
\end{align}
Note that in Equation (\ref{eq: n geo}), and in  subsequent calculations, we invoke the summation convention. 
It is also helpful to define the leading-order physical microscale unit normal $\bm{n}^{Y} = n^Y_i \bm{e}_i$ as follows
\begin{equation}
\label{eq: NY}
\bm{n}^{Y} = \frac{\displaystyle  \sigma_{ij} \frac{\partial f_s}{\partial y_j} \bm{e}_i}{\left[\displaystyle \sigma_{kl}\sigma_{km} \frac{\partial f_s}{\partial y_l}\frac{\partial f_s}{\partial y_m} \right]^{1/2}},
\end{equation}
such that $\bm{n}_s \sim \bm{n}^{Y}$ as $\epsilon \to 0$. Importantly, the transformed unit normal we work with in the homogenisation (Eq.~\ref{eq: NY}) is not equal to the geometric normal (Eq.~\ref{eq: n geo}) in general.
\end{subequations}

Thus using Equation (\ref{normals}), Equations (\ref{ad_diff}) become 
\begin{subequations}
\label{c_prob}
\begin{equation}\label{c_XY}
    \frac{\partial c}{\partial \tau}+\epsilon\frac{\partial {{c}}}{\partial {t}} = \left(\epsilon \frac{\partial}{\partial x_i}+\sigma_{ij} \frac{\partial}{\partial y_j}\right) \left[\frac{\partial c}{\partial x_i}+\frac{\sigma_{ik}}{\epsilon}\frac{\partial c}{\partial y_k}- \mathrm{Pe}_l \frac{{v_i}{c}}{\epsilon}\right], \quad {{\bm{y}}}\in\omega_f(x_1),
\end{equation}
\begin{multline}\label{Robin_2_scales}
    -\epsilon\gamma{c} \left[\displaystyle \sigma_{kl}\sigma_{km} n^y_ln^y_m \right]^{1/2} +\mathit{O}(\epsilon^2)= \left( \epsilon N_i+\sigma_{ij} n^y_j \right) \left[\frac{\partial c}{\partial x_i}+\dfrac{\sigma_{ik}}{\epsilon}\frac{\partial c}{\partial y_k}- \mathrm{Pe}_l \frac{{v_i}{c}}{\epsilon}\right], \quad {{\bm{y}}}\in \partial\omega_s(x_1), 
\end{multline}
with
\begin{equation}
    v_i, \quad c, \quad \text{periodic on} \quad \bm{y}\in\partial\omega_{=}(x_1)\ \ \text{and}\ \  \partial\omega_{||}(x_1),
\end{equation}
writing $\bm{v} = v_i \bm{e}_i$. Note that, for clarity of presentation in what follows, we have multiplied by $\epsilon$ when deriving Equation (\ref{c_XY}) from Equation~\eqref{c_hat}.
 \end{subequations}

\subsubsection{The dispersive homogenisation}
As for the flow problem, we consider an asymptotic expansion of the  concentration  field of the form 
\begin{multline}
\label{eq:c_expansion}
c(\bm{x}, \bm{y},t ,\tau ) = c^{(0)}(\bm{x}, \bm{y},t ,\tau ) + \epsilon c^{(1)}(\bm{x}, \bm{y},t ,\tau ) 
 + \epsilon^2 c^{(2)}(\bm{x}, \bm{y},t ,\tau ) +\cdots \quad \mbox{as} \quad \epsilon \to 0. 
\end{multline}

Equations (\ref{c_prob}) at leading order, $\mathcal{O}(1/\epsilon)$,  give
\begin{subequations}\label{ad_diff_leadingorder_Mo}
 \begin{equation}
0=  \sigma_{ij}\frac{\partial}{\partial y_j}\left(\sigma_{ik}\frac{\partial c^{(0)}}{\partial y_k}-\mathrm{Pe}_l\ v_i^{(0)}c^{(0)}\right), \ \quad {\bm{y}}\in{\omega}_f(x_1),
        \end{equation}
        \begin{equation}
       0= \sigma_{ij}n_j^y\left(\sigma_{ik}\frac{\partial c^{(0)}}{\partial y_k} -\mathrm{Pe}_l\ v_i^{(0)}c^{(0)}\right)
       , \quad {\bm{y}}\in \partial{\omega}_s(x_1), 
\end{equation}
and 
\begin{equation}
{v}_i^{(0)}, \quad c^{(0)}, \quad \text{periodic on} \quad \bm{y}\in\partial\omega_{=}
(x_1)\ \ \text{and}\ \  \partial\omega_{||}(x_1).
\end{equation}
\end{subequations}
As shown in Appendix \ref{Uniqueness}, 
the general solution to the system of Equations~(Eqs.~\ref{ad_diff_leadingorder_Mo}) is  that $c^{(0)}$ is independent of $\bm{y}$. This implies  $\partial c^{(0)}/\partial y_i\equiv0$, for $i=1,2$, and hence that $\langle c^{(0)}\rangle \equiv c^{(0)}$.

Proceeding to the next order, $\mathcal{O}(1)$, in Equation (\ref{c_prob}) yields
\begin{subequations}
\begin{align}
\label{dc0dtau}
    \frac{\partial c^{(0)}}{\partial \tau} &= 
    -\frac{\partial}{\partial x_i}\left(\mathrm{Pe}_l\ v_i^{(0)}c^{(0)}\right)+\sigma_{ij}\frac{\partial\mathcal{A}_i}{\partial y_j} ,\quad \bm{y}\in\omega_f,\\
     \label{2ndorderBC_Mo}
     0&=\sigma_{ij}n_j^y\mathcal{A}_i,\quad \bm{y}\in\partial\omega_s,
     \end{align}
     with
     \begin{equation}
     {v}_i^{(1)}, 
     c^{(1)}, \quad \text{periodic on} \quad \bm{y}\in\partial\omega_{=}(x_1)\ \ \text{and}\ \  \partial\omega_{||}(x_1),
   \end{equation}
   where 
   \begin{equation}
       \mathcal{A}_i \defeq \frac{\partial c^{(0)}}{\partial x_i}+\sigma_{ik}\frac{\partial c^{(1)}}{\partial y_k} - \mathrm{Pe}_l\left(v_i^{(0)}c^{(1)}+v_i^{(1)}c^{(0)}\right),
   \end{equation}
\end{subequations}
and where we have used $\partial c^{(0)}/\partial y_i = 0$ and the no-slip and no-penetration  boundary conditions on the solid surface  (Eq.~\ref{vi}).  
 Integrating Equation (\ref{dc0dtau}) over the transformed microscale fluid domain $\omega_f$ gives 

\begin{subequations}
\begin{equation}
\label{2ndorder_Mo}
   |\omega_f|\frac{\partial c^{(0)}}{\partial \tau} =  -\mathrm{Pe}_l\int_{\omega_f}\frac{\partial}{\partial x_i}\left({v}_i^{(0)}c^{(0)}\right) \mathrm{d}S_y + \int_{\omega_f} \frac{\partial\left(\sigma_{ij}\mathcal{A}_i\right)}{\partial y_j} \mathrm{d}S_y
\end{equation}
where we have noted that  $\sigma_{ij}$ is independent of $y_k$ for all $i, j, k =1,2$. On application of  the divergence theorem to the last term of  Equation~(\ref{2ndorder_Mo})  we find 
\begin{equation}
\label{junk}
    \int_{\omega_f} \frac{\partial\left(\sigma_{ij}\mathcal{A}_j\right)}{\partial y_j} \mathrm{d}S_y = \int_{\partial\omega_s} \sigma_{ij}\mathcal{A}_j n_j^y \mathrm{d}s_y  +\int_{\partial\omega} n^{\square}_j \sigma_{ij}\mathcal{A}_i\, \mathrm{d}s_y  \equiv 0
\end{equation}
where $\mathrm{d}s_y$ signifies an element of a scalar line integral, and $\bm{n}^{\square} = n^{\square}_i \bm{e}_i$ is the outward-facing unit normal to the external square boundary $\partial\omega$. Both terms on the right-hand side of  Equation (\ref{junk}) vanish;  the first term due to the boundary condition (Eq.~\ref{2ndorderBC_Mo})   and the second term as $\sigma_{ij}\mathcal{A}_i$ is periodic on $\partial\omega$.
\end{subequations}

Thus, applying the transport theorem (Eq.~\ref{transport_final}) to Equation (\ref{2ndorder_Mo}) leads to 
\begin{equation}
\label{pain_Mo}
\frac{\partial c^{(0)}}{\partial \tau} = -\frac{\mathrm{Pe}_l}{|\omega_f|}\left[\frac{\partial }{\partial x_i}\left(\int_{\omega_f}v_i^{(0)}c^{(0)}\, \mathrm{d}S_y  \right)+\int_{\partial\omega_s} \ N_iv_i^{(0)}c^{(0)}\  \mathrm{d}s_y\right],\quad \bm{y}\in\omega_f.
\end{equation}
The last term on the right-hand side of Equation (\ref{pain_Mo}) vanishes due to the no-slip and no-penetration condition (Eq.~\ref{vi}) on $\partial\omega_s$, hence 
Equation (\ref{pain_Mo}) becomes 
\begin{equation}
\label{Semi-final_Mo}
\frac{\partial c^{(0)}}{\partial \tau}=- \frac{\mathrm{Pe}_l}{\hat{\phi}}\bm{\nabla}_x\cdot\left(\hat{\phi}  \langle\bm{v}^{(0)}\rangle  c^{(0)}\right) = -\mathrm{Pe}_l  \langle\bm{v}^{(0)}\rangle  \cdot\bm{\nabla}_xc^{(0)},\quad \bm{y}\in\omega_f
\end{equation}
where we have use the definition of porosity (Eq.~\ref{phi}) and  macroscale incompressibility (Eq.~\ref{macro_incomp}; with $r=0$). Equation (\ref{Semi-final_Mo}) governs the leading-order 
fast-time evolution of concentration due to advection ---that is, the leading-order effect is that the solute pulse advects according to the fast time scale. This result is consistent with previous dispersion works (\textit{e.g.,} \cite{griffiths2013control}). However, 
recall that goal of this analysis remains to determine a macroscale equation for the concentration that balances all transport and removal mechanisms: advection, diffusion, dispersion and sorption. Since Equation (\ref{Semi-final_Mo}) does not yield any information on how the solute spreads as it is advected, we must continue until we determine an equation for how the concentration  varies relative to the slower timescale. 
To do this, we must proceed to a higher asymptotic order; the first step of this is to determine a closed system of equations for~$c^{(1)}$.

With this goal in mind we use the relationship determined in Equation (\ref{Semi-final_Mo}) to eliminate the fast-time derivative in  Equation (\ref{dc0dtau}), leading to 
\begin{multline}
\label{eq:fiddle_Mo}
  \sigma_{ij}\frac{\partial}{\partial y_j}\left(\frac{\partial c^{(0)}}{\partial x_i}+\sigma_{ik}\frac{\partial c^{(1)}}{\partial y_k}\right) =\\ 
  \mathrm{Pe}_l\left[-\langle v_i^{(0)}\rangle\frac{\partial c^{(0)}}{\partial x_i}+\frac{\partial }{\partial x_i}\left(v_i^{(0)}c^{(0)}\right)+\sigma_{ij}\frac{\partial}{\partial y_j}\left(v_i^{(0)}c^{(1)}+v_i^{(1)}c^{(0)}\right)\right] 
  ,\quad \bm{y}\in\omega_f.
\end{multline}

Using Equations (\ref{O0incomp}) and (\ref{Oiincomp}; with $r=0$) and that $c^{(0)}$ is independent of the microscale,  Equation (\ref{eq:fiddle_Mo}) becomes 
\begin{subequations}
\label{eq:c1_Mo}
\begin{multline}
    \sigma_{ij}\left[\mathrm{Pe}_l\ v_i^{(0)}\frac{\partial c^{(1)}}{\partial y_j}-\frac{\partial}{\partial y_j}\left(\frac{\partial c^{(0)}}{\partial x_i}+\sigma_{ik}\frac{\partial c^{(1)}}{\partial y_k}\right)\right] = \mathrm{Pe}_l\ \left(\langle v_i^{(0)}\rangle -v_i^{(0)}\right)\frac{\partial c^{(0)}}{\partial x_i}, \quad \bm{y}\in\omega_f,
\end{multline}
subject to 
\begin{equation}
    \sigma_{ij}n_j^y\left(\frac{\partial c^{(0)}}{\partial x_i}+\sigma_{ik}\frac{\partial c^{(1)}}{\partial y_k}\right)= 0, \quad \bm{y}\in\partial\omega_s.
\end{equation}
\end{subequations}
 
 The form of Equations \eqref{eq:c1_Mo} suggest that we can scale $\partial c^{(0)}/\partial x_i$ out of the problem via the substitution
\begin{equation}
\label{eq:Gamma_Mo}
c^{(1)}(\bm{x}, \bm{y},t ,\tau) = -\Gamma_n(\bm{x}, \bm{y})\frac{\partial}{\partial x_n}\Big[c^{(0)}(\bm{x},t ,\tau)\Big]+\breve{C}^{(1)}(\bm{x},t ,\tau),
\end{equation}
where  $\breve{C}^{(1)}$ is a scalar function and $\bm{\Gamma} = \Gamma_n\bm{e}_n$  is a vector function, such that $\langle\bm{\Gamma}\rangle=0$. 
Substituting  Equation \eqref{eq:Gamma_Mo} into  Equation \eqref{eq:c1_Mo}, we obtain a system of equations for $\Gamma_k$: 
\begin{subequations}
\label{Gamma_y}
\begin{align}
\sigma_{ij}\left[\mathrm{Pe}\ v_i^{(0)}\frac{\partial \Gamma_n}{\partial y_j} - \frac{\partial}{\partial y_j}\left(
\sigma_{ik}\frac{\partial \Gamma_n}{\partial y_k}\right)\right] = -\mathrm{Pe}_l\ \left(\langle v_n^{(0)}\rangle -v_n^{(0)}\right), \quad &\bm{y}\in\omega_f, \\
\sigma_{ij}n_j^y\left(\delta_{in}-\sigma_{ik}\frac{\partial \Gamma_n}{\partial y_k}\right)=0, \quad &\bm{y}\in\partial\omega_s\\
\Gamma_n \quad  \text{ periodic on} \quad &\bm{y}\in\partial\omega_{=}\ \ \text{and}\ \  \partial\omega_{||}.
\end{align}
\end{subequations}

Our overarching goal remains to derive the macroscale dispersion equation. To achieve this, we must consider one final asymptotic order.   Equation~(\ref{c_prob}) at $\mathcal{O}(\epsilon)$ gives
\begin{subequations}
\begin{align}
\label{eq:c_t_tau}
    \frac{\partial c^{(0)}}{\partial t}+    \frac{\partial c^{(1)}}{\partial \tau} =  \sigma_{ij}\frac{\partial \mathcal{B}_i}{\partial y_j} + \frac{\partial \mathcal{A}_i}{\partial x_i},& \quad \bm{y}\in\omega_f, \\
    \label{gammaBC}
    -\epsilon\gamma{c^{(0)}} \left[\displaystyle \sigma_{kl}\sigma_{km} n^y_ln^y_m \right]^{1/2} = \sigma_{ij}n_j^y\mathcal{B}_i + N_i\mathcal{A}_i, &\quad \bm{y}\in\partial\omega_s,
\end{align}
where
\begin{equation}
    \mathcal{B}_i \defeq \frac{\partial c^{(1)}}{\partial x_i}+\sigma_{ik}\frac{\partial c^{(2)}}{\partial y_k} - \mathrm{Pe}_l\ \left(v_i^{(2)}c^{(0)}+v_i^{(1)}c^{(1)}+v_i^{(0)}c^{(2)} \right).
\end{equation}
\end{subequations}

Integrating Equation (\ref{eq:c_t_tau}) over the transformed microscale fluid domain $\omega_f$ and using the transport theorem (Eq.~\ref{transport_final}) we obtain 
\begin{multline}
\label{cttau}
    |\omega_f|\frac{\partial c^{(0)}}{\partial t}+\frac{\partial }{\partial \tau}\left(\int_{\omega_f}c^{(1)}\ \mathrm{d}S_y\right) =   \int_{\omega_f}\sigma_{ij}\frac{\partial\mathcal{B}_i}{\partial y_j}\ \mathrm{d}S_y+ \frac{\partial}{\partial x_i}\int_{\omega_f}\mathcal{A}_i\ \mathrm{d}S_y+\int_{\partial\omega_s}N_i\mathcal{A}_i\ \mathrm{d}s_y.
\end{multline}
We deal with the first term  on the right-hand side of Equation (\ref{cttau}) by noting $\sigma_{ij}$ is independent of $\bm{y}$ and applying  the divergence theorem to $\sigma_{ij}\mathcal{B}_i$. Then, applying periodicity on the cell boundary and  Equation~(\ref{gammaBC}), we  determine that 
\begin{equation}
\label{38}
    \int_{\omega_f}\sigma_{ij}\frac{\partial\mathcal{B}_i}{\partial y_j}\ \mathrm{d}S_y = -\gamma{c^{(0)}} \int_{\partial\omega_s} \left[\displaystyle \sigma_{kl}\sigma_{km} n^y_ln^y_m \right]^{1/2}\mathrm{d}s_y - \int_{\partial\omega_s}N_i\mathcal{A}_i\ \mathrm{d}s_y. 
\end{equation}
Substituting Equation (\ref{38}) into Equation (\ref{cttau}), we obtain  
\begin{multline}
\label{something}
    |\omega_f|\frac{\partial c^{(0)}}{\partial t}+\frac{\partial }{\partial \tau}\left(\int_{\omega_f}c^{(1)}\ \mathrm{d}S_y\right) = \frac{\partial}{\partial x_i}\int_{\omega_f}\mathcal{A}_i\ \mathrm{d}S_y - \gamma{c^{(0)}} \int_{\partial\omega_s} \left[\displaystyle \sigma_{kl}\sigma_{km} n^y_ln^y_m \right]^{1/2}\mathrm{d}s_y. 
\end{multline}
\begin{subequations}
\label{dcdt_second_order}
Using Equation (\ref{eq:Gamma_Mo}) and the fact that $\langle \boldsymbol{\Gamma}\rangle \equiv 0$ by definition, Equation (\ref{something}) can be expressed as
\begin{multline}
    |\omega_f|\frac{\partial c^{(0)}}{\partial t}+\frac{\partial }{\partial \tau}\left(\int_{\omega_f}\breve{C}^{(1)}\ \mathrm{d}S_y\right) = \\ \bm{\nabla}_x\cdot\left\{\hat{\phi}\hat{\boldsymbol{D}}\cdot\bm{\nabla}_xc^{(0)}-\hat{\phi}\mathrm{Pe}_l\left[  \langle\bm{v}^{(0)}\rangle  \breve{C}^{(1)}+\left(\frac{1}{\hat{\phi}}\int_{\omega_f}\bm{v}^{(1)}\mathrm{d}S_y\right)c^{(0)}\right]\right\}-\hat{F}(\hat{\phi},\hat{a})c^{(0)},
\end{multline}
where
\begin{equation}
\label{source}
   \hat{F}(\hat{\phi},\hat{a})\defeq  \dfrac{\gamma}{|\omega_f|} \int_{\partial\omega_s} \!\left[\sigma_j \sigma_j n_j^y n_j^y \right]^{1/2} \, \mathrm{d}s_y,
   \end{equation}
   and where
\begin{equation}
\label{D_defn}
  \hat{\bm{D}}(\hat{\phi}, \hat{a}) \defeq \bm{I} - \hat{\bm{D}}_\text{obst} + \hat{\bm{D}}_\text{disp},
  \end{equation}
such that
\begin{align}
\label{Dobst_y}
\left(\hat{\bm{D}}_\text{obst}\right)_{ij} &\defeq \frac{1}{|\omega_f|}\int_{\omega_f}\sigma_{ik}\frac{\partial \Gamma_j}{\partial y_k} \mathrm{d}S_y, \\ 
\label{Ddisp_y}
\left(\hat{\bm{D}}_\text{disp}\right)_{ij}&\defeq -\frac{ \mathrm{Pe}_l}{ |\omega_f|} \left(\int_{\omega_f(x_1)} v_i^{(0)}{\Gamma_j} \mathrm{d}S_y\right).
\end{align}
\end{subequations}

We introduce the macroscale intrinsic averaged concentration and fluid velocity fields accurate up to and including the first-order asymptotic correction: $\hat{\mathcal{C}}$ and 
and $\hat{\bm{\mathcal{V}}}$, respectively.  In particular, $\hat{\mathcal{C}}$ is defined by
\begin{subequations}
\label{mathcals}
\begin{align}
\label{mathcalC}
\hat{ \mathcal{C}} \defeq & \  \langle c^{(0)}\rangle +\epsilon\langle c^{(1)} \rangle,\\  \equiv &  \  C^{(0)} - \epsilon \hat{\bm{\nabla}} C^{(0)}\cdot \langle \boldsymbol{\Gamma}\rangle+\epsilon\breve{C}^{(1)}, \\ \equiv   & \ C^{(0)} +\epsilon\breve{C}^{(1)},
\end{align} 
since $\boldsymbol{\Gamma}$ is constructed such that  $\langle \boldsymbol{\Gamma}\rangle \equiv 0$ and $c^{(0)}$ is independent of the microscale so that  $C^{(0)} \defeq \langle c^{(0)}\rangle  \equiv c^{(0)} $. 
Similarly, $\hat{\mathcal{V}}$ is defined 
\begin{equation}
    \label{mathcalV}
    \hat{\bm{\mathcal{V}}}\defeq \bm{V}^{(0)} + \epsilon\bm{V}^{(1)}, 
\end{equation} 
where 
\begin{equation}
\label{vdef}
\bm{V}^{(r)} \defeq  \langle\bm{v}^{(r)}\rangle
\end{equation}
for $r\in\{0,1\}$.
\end{subequations}
Note that $\hat{ \mathcal{C}}\sim C^{(0)}$ and $\hat{ \mathcal{V}}\sim V^{(0)}$. 

As previously discussed, the dispersive effects characterised via Equation (\ref{dcdt_second_order}) occur over a slower timescale than the advective effects characterised via Equation (\ref{Semi-final_Mo}). We can combine these into a single equation by collapsing our two timescales back into the physical single timescale using Equation (\ref{Eq_2ts})--- that is, we sum Equation (\ref{Semi-final_Mo}) and $\epsilon$ times Equation (\ref{dcdt_second_order}) and then recombine the timescales. 
This procedure yields
\begin{equation}
\label{eq:c_homogenised} 
\frac{\partial\hat{\mathcal{C}}}{\partial\hat{t}} = \frac{1}{\hat{\phi}}\hat{\bm{\nabla}}\cdot\left(\hat{\phi}\hat{\bm{D}}\cdot\hat{\bm{\nabla}}\hat{\mathcal{C}}-\hat{\phi}\mathrm{Pe}_g\bm{\hat{\mathcal{V}}}\hat{\mathcal{C}}\right)-\hat{F}(\hat{\phi},\hat{a})\hat{\mathcal{C}}, \quad \hat{\bm{x}}\in\hat{\Omega},
\end{equation}
where we have additionally  used the definition of Pe$_g$ (Eq.~\ref{Pes}), and have replaced $\bm{\nabla}_x$ with~$\hat{\bm{\nabla}}$  since Equation (\ref{eq:c_homogenised}) depends only on $\bm{x} = \hat{\bm{x}}$ and $\hat{t}$.    
Note that this rearrangement  can also be obtained  by considering the expansion of Equation (\ref{eq:c_homogenised}) subject to Equation (\ref{mathcals}), retaining leading-order and first-order corrections only.

\subsubsection{Transforming to the physical microscale coordinate}

To \hspace{2mm} interpret \hspace{2mm} the \hspace{2mm}  rate \\ \noindent $F(\hat{\phi},\hat{a})$ physically, it is helpful to map its definition Equation \eqref{source} to the physical microscale coordinate $\bm{Y}$, defined in Equation \eqref{Y_map}.  This integral coordinate transform is shown in detail in \S3.3.1 of  Auton \textit{et al}. \cite{auton2021homogenised} and leads to 
\begin{equation}
\label{eq: F transform}
    \hat{F}(\hat{\phi},\hat{a}) = \frac{\gamma|\partial\omega_s^\star|}{|\omega_f||\omega^\star|}\equiv \frac{\gamma|\partial\omega_s^\star|}{|\omega_f^\star|} \equiv \frac{\gamma|\partial\omega_s^\star|}{\hat{a}\hat{\phi}},
\end{equation}
where we have used Equations~(\ref{volume_int_y2Y}), (\ref{phi}) and~(\ref{normals}); $\hat{F}$  is just the product of the dimensionless adsorption rate and ratio of surface area of solid to fluid fraction in a transformed cell. 

Similarly, we map the physical microscale coordinate $\bm{Y}$, defined in Equation \eqref{Y_map}. This mapping transforms the cell  problem (Eqs.~\ref{Gamma_y}) to 
\begin{subequations}
\label{Gamma_Y_1}
\begin{align}
\label{uninspired1}
-\mathrm{Pe}_l\ v_i^{(0)^\star}\frac{\partial \Gamma^\star_n}{\partial Y_i} + \frac{\partial}{\partial Y_i}\left(
\frac{\partial \Gamma_n^\star}{\partial Y_i}\right) = \mathrm{Pe}_l\ \left(V_n^{(0)^\star}-v_n^{(0)^\star}\right), \quad &\bm{Y}\in\omega_f^\star, \\
\label{uninsipred2}
n_i^{Y^\star}\frac{\partial \Gamma^\star_n}{\partial Y_i}=n_n^{Y^\star}, \quad &\bm{Y}\in\partial\omega_s^\star\\
\label{uninsipred3}
\Gamma^\star_n \quad  \text{ periodic on} \quad &\bm{Y}\in\partial\omega_{=}^\star\ \ \text{and}\ \  \partial\omega_{||}^\star, 
\end{align}
with 
\begin{equation}
\label{uninsipred4}
   \hspace{2.5cm} \langle {\Gamma_n}^\star\rangle = {0}, 
    \end{equation}
\end{subequations}
such that Equations (\ref{Dobst_y}) and (\ref{Ddisp_y}) become
\begin{subequations}
\label{Ds}
\begin{equation}
\left(\hat{\bm{D}}_\text{obst}\right)_{ij} \defeq\frac{1}{|\omega_f^\star|}\int_{\omega_f^\star}\frac{\partial \Gamma_j^\star}{\partial Y_i} \mathrm{d}S_Y \equiv  \frac{1}{\hat{a}\hat{\phi}}\int_{\omega_f^\star}\frac{\partial \Gamma_j^\star}{\partial Y_i} \mathrm{d}S_Y
\end{equation}
and 
\begin{equation}
\label{Ddisp_Y_wo_K}
   \left(\hat{\bm{D}}_\text{disp}\right)_{ij}\defeq
   -\frac{ \mathrm{Pe}_l}{|\omega_f^\star|} \left(\int_{\omega_f^\star(x_1)} v_i^{(0)}{\Gamma_j} \mathrm{d}S_Y\right) 
   \equiv -\frac{ \mathrm{Pe}_l}{ \hat{a}\hat{\phi}} \left(\int_{\omega_f^\star(x_1)} v_i^{(0)}{\Gamma_j} \mathrm{d}S_Y\right). 
\end{equation}
\end{subequations}

To evaluate $\hat{\bm{D}}$, we solve the transformed  cell problem, (Eqs.~\ref{Gamma_Y}), numerically in COMSOL Multiphysics,  where $v_i^{(0)}$ is in general determined via the  solution to another cell problem, the details of which depend on the prescribed flow. In \S\ref{exs}, we prescribe Stokes flow and specify the appropriate cell problems. While Stokes flow is typically associated with slow flow, and dispersion with fast flow, we note that both are consistent when $\tilde{D}\ll \tilde{V}\tilde{h}\ll\tilde{\nu}$, where $\tilde{\nu}$ is the kinematic viscosity of the fluid;   this gives the local Reynolds number $ \mathrm{Re}_l\defeq\tilde{V}\tilde{L}/\tilde{\nu}\ll1$ and $\mathrm{Pe}_l\equiv\tilde{V}\tilde{h}/\tilde{D}\gg1$. 

Using the definition of $\hat{\bm{\mathcal{V}}}$ (Eq.~\ref{mathcalV}) and the incompressibility of $\bm{v}^{(0)}$ and  $\bm{v}^{(1)}$ (Eq.~\ref{macro_incomp}), we find that $\hat{\bm{\mathcal{V}}}$ is also incompressible --- that is, 
\begin{equation}
    \hat{\bm{\nabla}}\cdot \hat{\bm{\mathcal{V}}} = 0
\end{equation}
Hence, Equation \eqref{eq:c_homogenised} becomes 
\begin{subequations}
\label{dispersion_equation}
\begin{equation}
\label{C_pre_drift} 
\frac{\partial\hat{\mathcal{C}}}{\partial\hat{t}} = \frac{1}{\hat{\phi}}\hat{\bm{\nabla}}\cdot\left(\hat{\phi}\hat{\bm{D}}        \cdot\hat{\bm{\nabla}}\hat{\mathcal{C}}\right)-\mathrm{Pe}_g\bm{\hat{\mathcal{V}}}\cdot\hat{\bm{\nabla}}\hat{\mathcal{C}}-\gamma\frac{|\partial\omega_s^\star|}{\hat{a}\hat{\phi}}\hat{\mathcal{C}}, \quad \hat{\bm{x}}\in\hat{\Omega},
\end{equation}
with 
\begin{equation}
\label{Dhatdefn}
    \hat{\bm{D}} = \bm{I} - \hat{\bm{D}}_\text{obst} + \hat{\bm{D}}_\text{disp}
\end{equation}
such that  $ \hat{\bm{D}}_\text{obst}$ and $\hat{\bm{D}}_\text{disp}$ are defined in Equation (\ref{Ds}). 
\end{subequations}
Recalling that $\mathrm{Pe}_g=\mathcal{O}(1/\epsilon)$, it might appear as though the second term on the right-hand side of Equation  (\ref{C_pre_drift}) is asymptotically unbalanced.  This is a consequence of the advective effects being important over a faster timescale than the dispersive effects, as also occurs in classic dispersion analyses. To understand the leading-order importance of the dispersive effects we now perform a drift transformation into the advective frame.

\subsection{Rescaled Dispersion Equation}

In this section we rescale the dispersion equation (Eq.~\ref{dispersion_equation}) to derive a leading-order equation for the removal, diffusion and dispersion of a solute. As previously mentioned, the leading-order effect of the flow on the solute is the solute pulse advects with the flow. Here, we separate these effects ---   
 we take the leading-order velocity of the advection of the solute pulse and then investigate the leading order spreading around the pulse. 
Since the fast flow is governed by the global P\'{e}clet number such that $\mathrm{Pe}_g = \mathcal{O}(1/\epsilon)$, we first rescale the spatial domain via 
 \begin{subequations}
 \begin{equation}
     \bm{\hat{x}} = \frac{\bm{X}}{\epsilon}
 \end{equation} 
 so that Equation (\ref{C_pre_drift}) becomes
\begin{equation}
\label{Another_C_eq}
         \frac{\partial\mathcal{C}}{\partial \hat{t}} - \frac{\epsilon^2}{\phi}\bm{\nabla}_X\cdot\left(\phi\bm{D}\cdot\bm{\nabla}_X\mathcal{C}\right)+\mathrm{Pe}_l\left(\bm{\mathcal{V}}\cdot\bm{\nabla}_X\right)\mathcal{C} = -\gamma\frac{|\partial\omega_s^\star|}{a\phi}\mathcal{C},
\end{equation}
where we denote functions of $\bm{X}$  (independent of $\hat{\bm{x}}$) without a hat. 
 \end{subequations}

To investigate the diffusive and dispersive solute transport, we then rescale around   a pulse of solute, which advects with the flow:
\begin{subequations}
\label{drift}
\begin{align}
    \bm{X} &= \bm{X}_0(T)+\epsilon\bm{\xi},\\
    \hat{t} &= T,\\
    \frac{\mathrm{d} \bm{X}_0}{\mathrm{d}T} &= \mathrm{Pe}_l\bm{\mathcal{V}}(\bm{X}_0) = \mathrm{Pe}_l\left(\bm{{V}}^{(0)}(\bm{X}_0) + \epsilon\bm{{V}}^{(1)}(\bm{X}_0)\right),
\end{align} 
with 
\begin{equation}
    \bm{X}_0(0) = 0,
\end{equation}
 so that 
\begin{multline}
   \bm{\nabla}_\xi = \epsilon\bm{\nabla}_X
   \quad\text{and}\quad  \frac{\partial}{\partial T} = \frac{\partial}{\partial \hat{t}} + \mathrm{Pe}_l\bm{\mathcal{V}}(\bm{X}_0)\cdot\bm{\nabla}_X = \frac{\partial}{\partial \hat{t}} + \mathrm{Pe}_l\left(\frac{\bm{{V}}^{(0)}(\bm{X}_0)}{\epsilon}+\bm{{V}}^{(1)}(\bm{X}_0)\right)\cdot\bm{\nabla}_\xi. 
\end{multline}
\end{subequations}
\begin{figure}
    \centering
    \includegraphics[width=\textwidth]{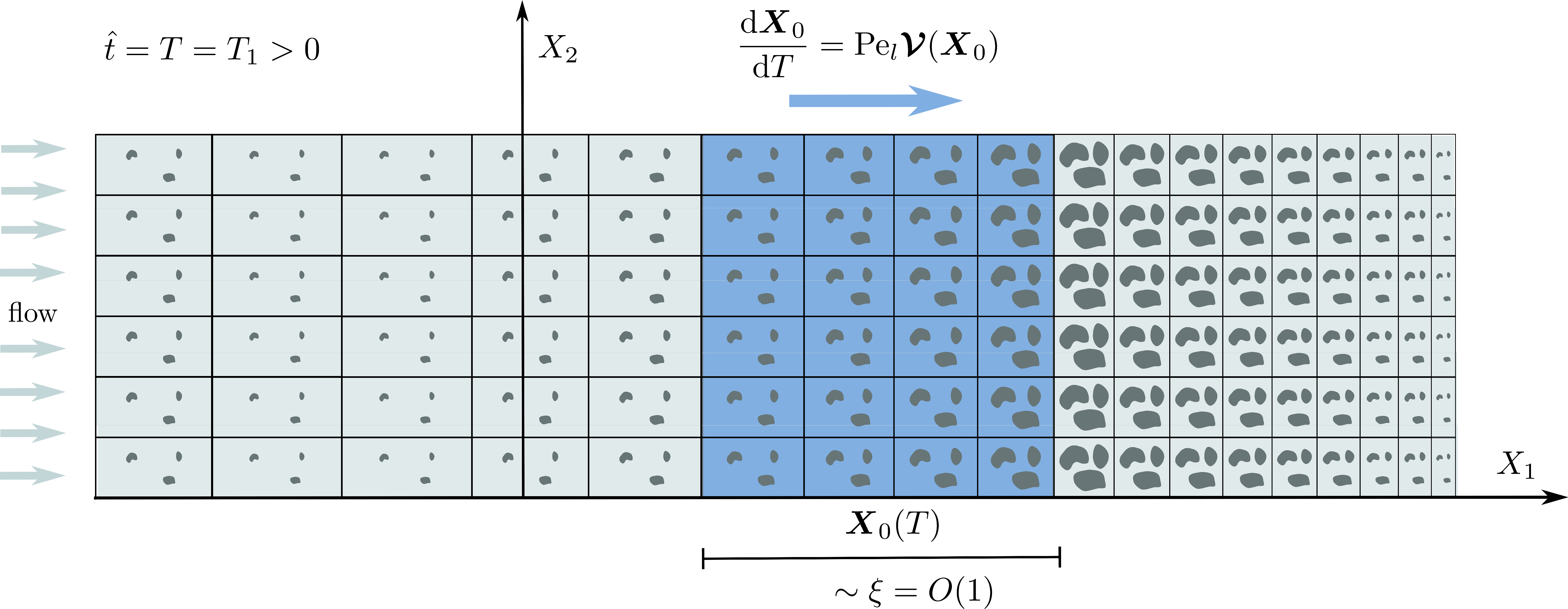}
    \caption{    We consider the evolution of a pulse of solute at some later time $\hat{t} =T=T_1$ subject to fast fluid flow ($\mathrm{Pe}_g=\mathcal{O}(1/\epsilon)$). Advection dominates the propagation of the solute and thus to see the diffusive and dispersive effects, we define a drift transformation (Eqs.~\ref{drift}) that advects with the flow, the speed of which is shown on the schematic. The centre of the pulse is given by the dimensionless co-ordinate $X_0(T)$ such that $X_0(0)=0$ and the $\bm{\xi}$-axis moves with the pulse enabling investigating relative to the dominant advective transport. Note that $\bm{\xi}$ is scaled so that the solute pulse is $\mathcal{O}(1)$ relative to $\bm{\xi}$. 
      \label{fig:schem_dim_drift} }
\end{figure}
This transformation places the origin of the  $\bm{\xi}$-axes at the centre of the solute pulse, which  travels  with velocity $\mathrm{Pe}_l\bm{\mathcal{V}}(\bm{X}_0)$ (see Figure \ref{fig:schem_dim_drift}). This allows us to understand the evolution of the spread of the solute from its initial pulse of unit dimensionless length (relative to $\bm{\xi}$).
Under this scaling, $\mathcal{O}(1)$ variations in the solute concentration will be characterised by  $\mathcal{O}(1)$ variations in $\bm{\xi}$; thus we enforce \begin{equation}
    {\mathcal{C}}\left(\bm{X},\hat{t}\right) = \mathcal{C}(\bm{\xi},T).
\end{equation}
As the solute pulse propagates along the length of the filter ($\bm{X}$), the microstructure of the filter, characterised by $\phi$ and $a$, varies. This variation tells us that $\bm{\mathcal{{V}}} = \bm{\mathcal{V}}(\bm{X}$), $\bm{D} = \bm{D}(\bm{X})$, $\phi=\phi(\bm{X}) $
and $a = a(\bm{X})$
which are locally given by
\begin{subequations}
\label{Taylors}
\begin{align}
\label{TaylorV}
     \bm{\mathcal{V}}(\bm{X}=\bm{X}_0+\epsilon\bm{\xi}) &= \bm{\mathcal{V}}(\bm{X}_0)+\epsilon\left(\bm{\xi}\cdot\bm{\nabla}_X\right)\left.\bm{\mathcal{V}}\right|_{\bm{X}=\bm{X}_0}+\cdots\\
     \bm{D}(\bm{X}=\bm{X}_0+\epsilon\bm{\xi},T) &= \bm{D}(\bm{X}_0,T)+\epsilon\left(\bm{\xi}\cdot\bm{\nabla}_X\right)\left.\bm{D}\right|_{\bm{X}=\bm{X}_0}+\cdots\\
          \phi(\bm{X}=\bm{X}_0+\epsilon\bm{\xi},T) &= \phi(\bm{X}_0,T)+\epsilon\left(\bm{\xi}\cdot\bm{\nabla}_X\right)\left.\phi\right|_{\bm{X}=\bm{X}_0}+\cdots.
\end{align}
\end{subequations}
Using Equations (\ref{drift}) and  (\ref{TaylorV}), Equation (\ref{Another_C_eq}) becomes 
\begin{equation}
\label{MMS_then_drift}
            \frac{\partial\mathcal{C}}{\partial T} = \left.(\bm{D})_{ij}\right|_{\bm{X}=\bm{X}_0}\frac{\partial^2\mathcal{C}}{\partial\xi_i\partial\xi_j}-\mathrm{Pe}_l\xi_i \frac{\partial \mathcal{C}}{\partial \xi_j}\left.\frac{\partial\mathcal{V}_j }{\partial X_i} \right|_{\bm{X}=\bm{X}_0} -\gamma\frac{|\partial\omega_s^\star|}{a\phi}\mathcal{C},
\end{equation}
where the drift transformation accounts for  the leading-order advective transport of $\mathcal{C}$. Note that the second term on the right-hand side of Equation (\ref{MMS_then_drift}) gives an $\mathcal{O}(1)$ correction for advection. 

Finally,  although there is no explicit dependence on $\phi$ or $a$ except in the removal term, the position $\bm{X}_0$ travels through the porous media and thus the effects of the varying microstructure are reflected in the effective diffusivity and flow gradients.

\section{Illustrative Example}
\label{exs}

In this section, we prescribe a Stokes flow and examine a specific pore structure  where the solid domain constitutes an array of solid circular obstacles in a hexagonal array. This could model, for example, the microscale geometry of a granular material. 

\subsection{Stokes flow} Dimensionless Stokes flow is given by 
\begin{equation}
-\hat{\bm{\nabla}}\hat{p}+\epsilon^2\hat{\nabla}^2\hat{\bm{v}}=\bm{0}
\end{equation}
where $\hat{p}$ is the dimensionless pressure,  which has been  scaled to balance macroscopic pressure gradient with
viscous dissipation at the pore scale. 
The full homogenisation of Stokes flow  for a material subject to the same two degrees of microstructural freedom is given in Auton \textit{et al}.~\cite{auton2021homogenised}. Here, for completeness, we briefly recap the relevant results of the flow homogenisation in Auton \textit{et al}.~\cite{auton2021homogenised}. In particular, we present the flow cell-problem which we subsequently solve  numerically using COMSOL multiphysics. The flow homogenisation leads to a set of equations for the microscale velocity  and pressure: 
\begin{subequations}
\begin{align}
\label{Darcy_micro}
\bm{v}^{(0)} &= -\bm{{\mathcal{K}}}(\bm{x}, \bm{y}) \cdot
\boldsymbol{\nabla}_x p^{(0)}, \qquad \bm{y}\in\omega_f, 
\\
p^{(1)} &= -\bm{\Pi}(\bm{x},\bm{y})\cdot\boldsymbol{\nabla}_x p^{(0)} +\breve{p}(\bm{x}),\qquad \bm{y}\in\omega_f,
\end{align}
\end{subequations}
where $p^{(i)}$ is the $i^\text{th}$-order microscale  pressure, $\breve{p}(\bm{x})$ is a scalar function which remains undetermined but is not important to our analysis, and where  $\bm{{\mathcal{K}}}(\bm{x}, \bm{y})$ is a tensor function  and $\bm{\Pi}(\bm{x}, \bm{y})$ is a vector function, both of which are numerically determined via solution of the flow cell-problem 
\begin{subequations}
\label{stokes_homog_O_ep_scaled}
    \begin{align}
        \bm{I}-\boldsymbol{\nabla}_y^a\otimes\bm{\Pi}+\left(\nabla_Y^a\right)^2\bm{{\mathcal{K}}}= \bm{0}, &\quad {\bm{y}}\in\omega_f(x_1), \label{stokes_main_O_ep_scaled} \\
    \boldsymbol{\nabla}_y^a\cdot{\bm{{\mathcal{K}}}} = \bm{0}, &\quad {\bm{y}}\in\omega_f(x_1), \label{incomp_O_ep_scaled} \\
        \bm{{\mathcal{K}}} = \bm{0}, &\quad {\bm{y}}\in \partial\omega_s(x_1),
\end{align}
with 
\begin{equation}
\mathcal{K}_{ij}\defeq\left(\bm{\mathcal{K}}\right)_{ij} \ \  \text{and} \ \  \Pi_{i}\defeq\left(\bm{\Pi}\right)_{i} \quad  \text{periodic on} \quad \bm{y}\in\partial\omega_{=}  \ \  \text{and}\ \  \partial\omega_{||},  
\end{equation}
 where
\begin{equation}
\label{eq: tensor def}\left(\boldsymbol{\nabla}_y^a\otimes\bm{\Pi}\right)_{ij} = \frac{\partial \Pi_j}{\partial y_i} \quad  \text{and} \quad(\boldsymbol{\nabla}_y^a\cdot{\bm{{\mathcal{K}}}})_{i} =  \frac{\partial {\mathcal{K}}_{ji}}{\partial y_j}.
\end{equation}
\end{subequations}

In the physical microscale coordinate $\bm{Y}$, Equations (\ref{stokes_homog_O_ep_scaled})  becomes
\begin{subequations}
\label{stokes_homog_O_ep_scaled_naive}
    \begin{align}
        \bm{I}-\boldsymbol{\nabla}_Y\otimes\bm{\Pi}^\star+\nabla_Y^2\bm{{\mathcal{K}}}^\star = \bm{0}, &\quad {\bm{Y}}\in\omega_f^{\star}(x_1), \label{stokes_main_O_ep_scaled_naive} \\
        \boldsymbol{\nabla}_Y\cdot{\bm{{\mathcal{K}}}}^\star = \bm{0}, &\quad {\bm{Y}}\in\omega_f^{\star}(x_1), \label{incomp_O_ep_scaled_naive} \\
        \bm{{\mathcal{K}}}^\star = \bm{0}, &\quad {\bm{Y}}\in \partial\omega_s^{\star}(x_1),
\end{align}
with 
\begin{equation}    \mathcal{K}^\star_{ij}\defeq\left(\bm{\mathcal{K}}^\star\right)_{ij} \ \  \text{and} \ \  \Pi_{i}^\star\defeq\left(\bm{\Pi}^\star\right)_{i} \quad  \text{periodic on} \quad \bm{Y}\in\partial\omega_{=}^{\star}  \ \  \text{and}\ \  \partial\omega_{||}^{\star}. 
\end{equation}
\end{subequations}

On taking the intrinsic  average of  Equation (\ref{Darcy_micro}),  we determine the homogenised  flow equation, essentially equivalent to Darcy's equation:

\begin{equation}
\label{Stokes_min}
\bm{V}^{(0)} = -\langle\bm{{\mathcal{K}^\star}}(\bm{x}, \bm{Y})\rangle \cdot
\boldsymbol{\nabla}_x \hat{P},
\end{equation}
where $\hat{P}(\hat{x}_1)\defeq\langle p^{(0)}\rangle$ is the macroscale leading-order pressure, and 
the macroscale permeability tensor \begin{equation}
    \hat{\bm{K}} \defeq \langle\bm{{\mathcal{K}^\star}}(\bm{x}, \bm{Y})\rangle\equiv\langle\bm{{\mathcal{K}}}(\bm{x}, \bm{y})\rangle
    \end{equation}
    is the intrinsic average of $\bm{{\mathcal{K}^\star}}$ over the physical microscale, or equivalently is the intrinsic average of $\bm{{\mathcal{K}}}$ over the transformed microscale.

Using the velocity decomposition given in Equation~(\ref{Darcy_micro}), the transport cell-problem  (Eqs.~\ref{Gamma_Y_1}) 
becomes 
\begin{subequations}
\label{Gamma_Y}
\begin{multline}
\label{Gamma_1}
-\mathrm{Pe}_l\mathcal{K}_{ij}^\star\frac{\partial \hat{P}}{\partial x_j}\frac{\partial \Gamma^\star_n}{\partial Y_i} - \frac{\partial}{\partial Y_i}\left(
\frac{\partial \Gamma_n^\star}{\partial Y_i}\right) =  -\mathrm{Pe}_l\ \left(\frac{1}{|\omega_f^\star|}\int_{\omega_f^\star}-\mathcal{K}_{nj}^\star\ \mathrm{d}S_Y+\mathcal{K}_{nj}^\star \right)\frac{\partial \hat{P}}{\partial x_j}, \quad \bm{Y}\in\omega_f^\star,
\end{multline}
\begin{align}
\hspace{3cm} &n_i^{Y^\star}\frac{\partial \Gamma^\star_n}{\partial Y_i}=n_n^{Y^\star}, \quad \bm{Y}\in\partial\omega_s^\star\\
&\Gamma^\star_n  \quad  \text{periodic on} \quad \bm{Y}\in\partial\omega_{=}^\star\ \ \text{and}\ \  \partial\omega_{||}^\star, 
\end{align}
with \begin{equation}
\label{Gamma_cond}
    \langle {\Gamma_n}^\star\rangle = {0},
    \end{equation}
and Equation~(\ref{Ddisp_Y_wo_K}) becomes 
\begin{equation}
\label{Ddisp_final}
   \left(\hat{\bm{D}}_\text{disp}\right)_{ij}\defeq -\frac{ \mathrm{Pe}}{ \hat{a}\hat{\phi}} \left(\int_{\omega_f^\star(x_1)} \mathcal{K}_{ik}{\Gamma_j} \mathrm{d}S_Y\right) \frac{\partial \hat{P}}{\partial x_k}. 
\end{equation}
\end{subequations}

\subsection{Filter geometry}

\begin{figure}
    \centering
    \includegraphics[width=\textwidth]{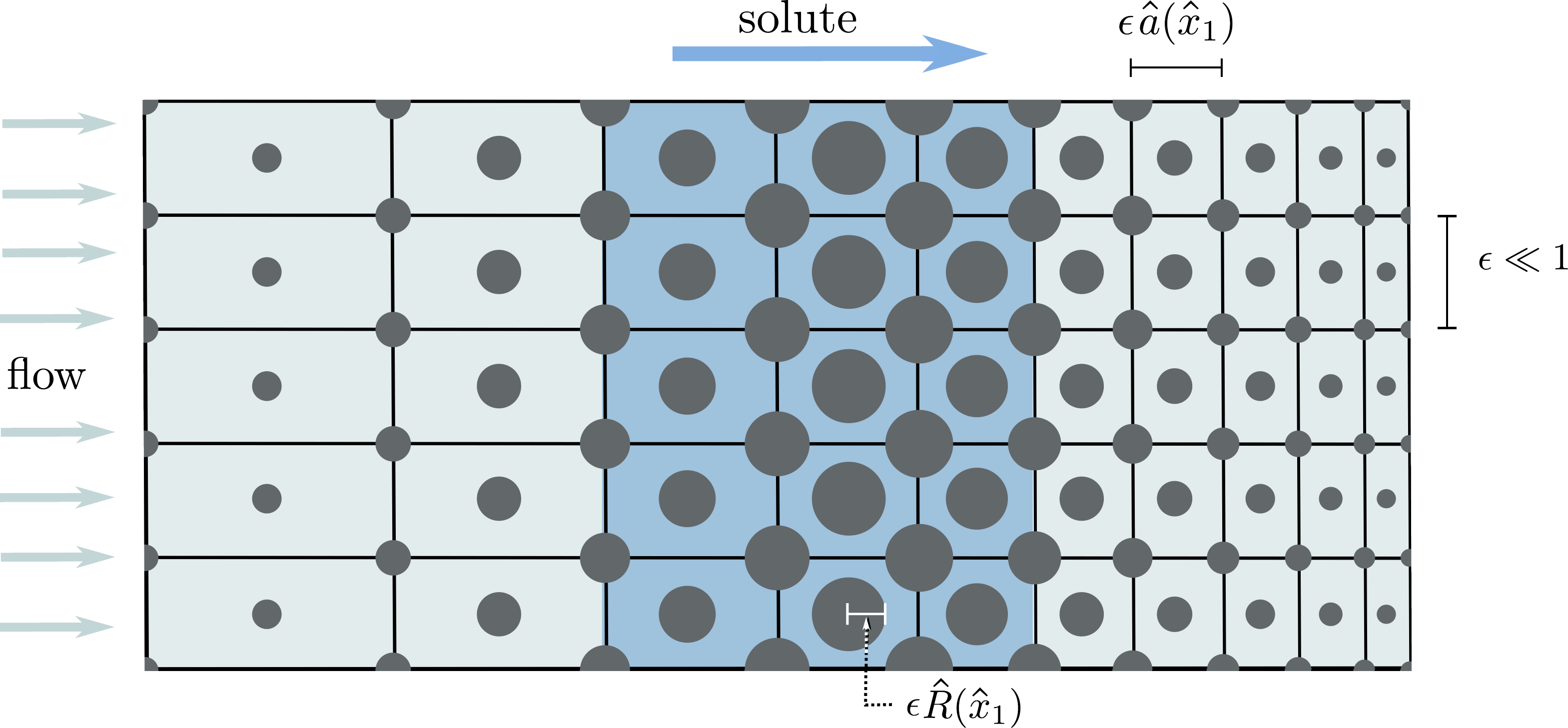}
    \caption{ \label{fig:schem_dim} We consider the flow of fluid carrying solute through a heterogeneous porous material in two dimensions for a specific illustrative example. Here, the porous medium is formed of an array of circular obstacles of dimensionless radius $\hat{R}(\hat{x}_1)$,  located in both the centre and at the corners of a rectangular cell of transverse height $\epsilon\ll1$ and longitudinal width $\epsilon \hat{a}(\hat{x}_1)$.}
\end{figure}

\begin{figure}
    \centering
    \includegraphics[width=1\textwidth]{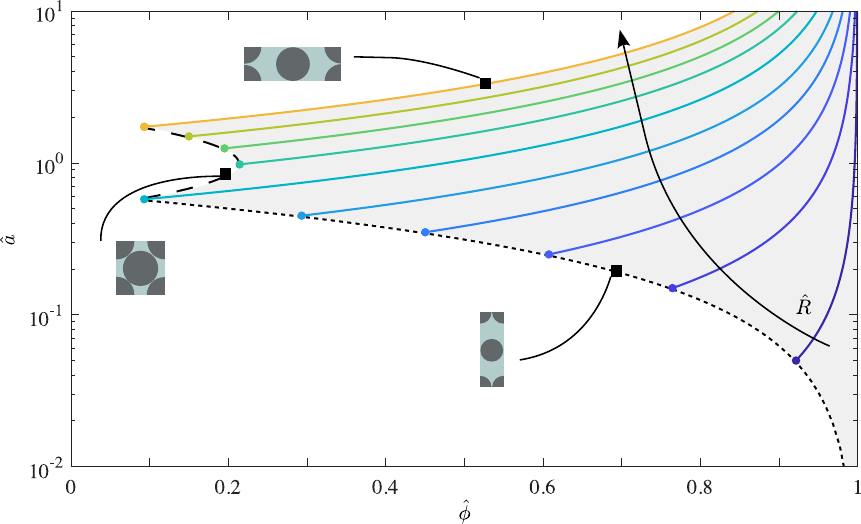}
    \caption{The porosity $\hat{\phi}$ of a hexagonal cell increases with aspect ratio $\hat{a}$ and decreases with obstacle radius $\hat{R}$ according to Equation~\eqref{phi_a_R}. We show $\hat{a}$ versus $\hat{\phi}$ for \mbox{$\hat{R}\in\{0.025, 0.075, 0.125, 0.175, 0.225, 1/(2\sqrt{3}), 0.35,0.4, 0.45, 0.5\}$} (dark to light). The attainable region of the $\hat{\phi}$-$\hat{a}$ plane (shaded grey)  is defined by $\hat{a}\geq \hat{a}_\text{min}(\hat{R})$ (Eq.~\ref{amin}), for $0<\hat{R}\leq1/2$ with the $\hat{\phi}$--$\hat{a}$--$\hat{R}$ relationship defined in Equation~(\ref{phi_a_R}). 
    In particular,  when  $0<\hat{R}< \hat{R}_\text{crit}$, $\hat{a}_\text{min} \equiv 2\hat{R}$ so that the lower bound (dotted line) is defined by $\hat{\phi} = 1-\pi \hat{a}/2$, with $\hat{a}\in(0,2\hat{R}_\text{crit})$; along this boundary we lose transverse connectivity but maintain longitudinal connectivity. When $\hat{R}_\text{crit}\leq \hat{R}<1/2$, $\hat{a}_\text{min} \equiv\sqrt{16\hat{R}^2-1}$ so that  the left-hand bound (dashed line) is defined by $\hat{\phi} = 1-\pi(\hat{a}^2+1)/(8\hat{a})$, with $\hat{a}\in[1/\sqrt{3}, \sqrt{3})$;  along this boundary we lose  connectivity in both the longitudinal and transverse directions. The upper bound (yellow line) is attained when $\hat{R} = 1/2$ and is parameterised by $\hat{\phi} = 1-\pi/(2\hat{a})$, for $\hat{a}\in[\sqrt{3},\infty)$; along this boundary we lose longitudinal connectivity but retain transverse connectivity. 
    Note that the smallest attainable $\hat{\phi}$ for any $\hat{R}$, $\hat{a}$ combination is  $\hat{\phi}_\text{min}=1-\pi/(2\sqrt{3})$ which is achieved for with two distinct combinations of $\hat{a}$ and $\hat{R}$:  $\hat{a}= 1/\sqrt{3}$, $\hat{R}=1/(2\sqrt{3})$ and $\hat{a} = \sqrt{3}$, $\hat{R}=1/2$. \label{fig:AR_phi} 
}
\end{figure}

 Here, we consider a specific pore structure  which has a  solid domain comprising solid circular obstacles in a hexagonal array. Specifically, each rectangular cell contains a fixed, rigid circular obstacle of dimensionless radius $\hat{R}(\hat{x}_1)$ at its centre and quarter circles of radius $\hat{R}(\hat{x}_1)$ at each corner (see Figure \ref{fig:schem_dim}). Since $\hat{R}(\hat{x}_1)$ controls the obstacle size over the length of the medium, we take the scale factor $\hat{\lambda}(\hat{x}_1) = \hat{R}(\hat{x}_1)$. 
 The height of each cell is fixed, thus, the maximum possible value of $\hat{R}$ is $1/2$.  
To prevent the obstacles from overlapping, we must enforce conditions on $\hat{a}$ for a given $\hat{R}$--- that is, we define a minimum value of $\hat{a}$ that depends on the value of $\hat{R}$:
\begin{subequations}
\begin{equation}
\label{amin}
   \hat{a}_\text{min}(\hat{R}) \defeq
\left\{
    \begin{array}{lr}
        \hat{a}=2\hat{R}, & \quad\text{if} \quad0<\hat{R}< \hat{R}_\text{crit} \\
        \hat{a}=\sqrt{16\hat{R}^2-1}, & \quad \text{if} \quad \hat{R}_\text{crit}\leq \hat{R} \leq1/2.
    \end{array}
\right. 
\end{equation}
where 
\begin{equation}
\label{Rcrit}
\hat{R}_\text{crit}\defeq\frac{1}{2\sqrt{3}}, 
\end{equation}
and with corresponding minimum porosity $\hat{\phi}_\text{min}(\hat{R})$.
\end{subequations}
We enforce that $\hat{a}\geq \hat{a}_\text{min}(\hat{R})$ and, consequently, $\phi\geq\hat{\phi}_\text{min}(\hat{R})$.
When $0<\hat{R}\leq \hat{R}_\text{crit}$ and $\hat{a} = \hat{a}_\text{min}(\hat{R})$ we lose transverse connectivity of the domain (Figure \ref{fig:AR_phi}; dotted line). When $\hat{R}_\text{crit}\leq \hat{R} \leq 1/2$  and $\hat{a} = \hat{a}_\text{min}(R)$   we lose both the  transverse and longitudinal  directions (Figure \ref{fig:AR_phi}; dashed line). Further, when $\hat{R}=1/2$ and $\hat{a}= \hat{a}_\text{min}(R)\equiv\sqrt{3}$ then there is connectivity in the transverse direction but no connectivity in the longitudinal direction (Figure \ref{fig:AR_phi}; yellow line).   The boundary on the right of the domain corresponds to the limit $\hat{\phi}\to1$, which  occurs when $\hat{R}\to0$ and when  $\hat{a}\to\infty$.  These constraints define the attainable region of the  $\hat{a}$, $\hat{R}$ and $\hat{\phi}$ parameter space (Figure~\ref{fig:AR_phi}; shaded grey). 

This construction leads to a porous medium whose properties vary in the longitudinal direction but not in the transverse direction (Figure~\ref{fig:schem_dim}). The microstructure depends on $\hat{a}$, $\hat{\phi}$ and $\hat{R}$, any two of which are independent and the third prescribed by the geometric relation
\begin{equation}
\label{phi_a_R}
\hat{\phi}(\hat{x}_1) = \frac{|\omega_f(\hat{x}_1)|}{|\omega(\hat{x}_1)|} = \frac{|\omega_f^{\star}(\hat{x}_1)|}{|\omega^{\star}(\hat{x}_1)|} \equiv
1-\frac{2\pi \hat{R}(\hat{x}_1)^2}{\hat{a}(\hat{x}_1)}, 
\end{equation}
where we have used $|\omega^{\star}| = \hat{a}$ and $|\omega_f^{\star}| = \hat{a} - 2\pi \hat{R}^2$. 

For this geometry, we may explicitly evaluate the effective adsorption rate $\hat{F}(\hat{\phi},\hat{a})$ in Equation~\eqref{source} using the formulation from Equation~\eqref{eq: F transform}, giving
\begin{equation}
\label{source_rect}
\hat{F}(\hat{\phi},\hat{a}) = \dfrac{\gamma|\partial \omega^{\star}_s|}{|\omega^{\star}_f|} = \frac{\gamma|\partial\omega_s^\star|}{\hat{a}\hat{\phi}}= \frac{4 \gamma\pi \hat{R}}{\hat{a} \hat{\phi} } = \frac{2\gamma\left(1-\hat{\phi}\right)}{\hat{R} \hat{\phi}}. 
\end{equation}

\begin{figure}
    \centering
        \includegraphics[width=1\textwidth]{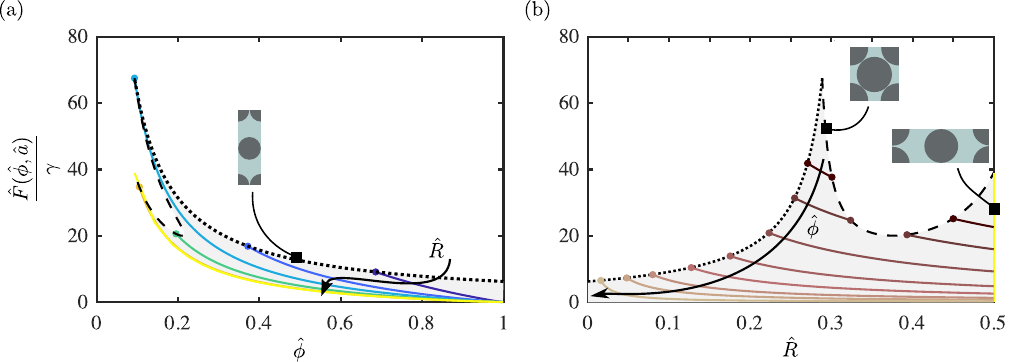}
    \caption{\label{F} The effective adsorption rate $\hat{F}$ normalised with $\gamma$   against (a) $\phi$  for $\hat{R}\in\{0.1, 0.2, \hat{R}_\text{crit}, 0.4, 0.49\}$ (blue to yellow) and (b) $\hat{R}$ for fixed values of  $\hat{\phi}\in\{0.15, 0.2, 0.3, 0.45, 0.6, 0.75, 0.85, 0.95\}$ (dark to light). The attainable region of the $\hat{\phi}$-$\hat{F}/\gamma$ plane (shaded grey) is bounded by the constraints that $\hat{a}\geq \hat{a}_\text{min}$ for $0<\hat{R}\leq1/2$, with the line styles of the boundary the same as in Figure~\ref{fig:AR_phi}.     
    In all cases, $\hat{F}$ is as defined in Equation~(\ref{source_rect}).}
\end{figure}

In Figure~\ref{F}, we investigate the effect of microscale geometry on the macroscale solute removal via a numerical analysis of Equation~(\ref{source_rect}).  
The partially absorbing boundary condition on the microscale (\textit{cf.} Equation~(\ref{Robin_dim})) leads to an effective sink term $\hat{F}$ in the macroscale transport problem given by Equation~(\ref{source_rect});  $\hat{F}$ is the product of  the dimensionless adsorption rate $\gamma$ and the ratio of the perimeter of an obstacle ($4\pi \hat{R}$) to the fluid area within a cell  ($\hat{a}\hat{\phi}$) for a hexagonal array of circular obstacles. Hence, the strength of the sink term is proportional to $\gamma$ and depends strongly on the microscale geometry of the problem. We consider $\hat{F}/\gamma$ to isolate the effects of microstructure. The attainable  region of
the $\hat{\phi}$-$\hat{F}/\gamma$ plane (Figure~\ref{F}, left; shaded grey) is bounded by  the constraints  $\hat{a}\geq \hat{a}_\text{min}$ and $0<\hat{R}\leq1/2$; the boundaries have the same line styles and colours as the corresponding boundaries in Figure~\ref{fig:AR_phi}.  Fixing $\hat{R}\in\{0.1, 0.2, R_\text{crit}, 0.4, 0.49\}$ (Figure~\ref{F}a) where $\hat{R}_\text{crit}$ is defined in Equation~(\ref{Rcrit}), we see  
 that the maximum removal for a given $\hat{R}$ is achieved when $\hat{a}=\hat{a}_\text{min}$. Further, the global maximal removal is achieved for $\hat{R}=\hat{R}_\text{crit}$ and $a_\text{min}=2R_\text{crit}$ 
Fixing $\hat{\phi}\in\{0.15, 0.2, 0.3, 0.45, 0.6, 0.75, 0.85, 0.95\}$ (Figure~\ref{F}b) we see that the maximum removal for a given $\hat{\phi}$ is also achieved when $\hat{a}=\hat{a}_\text{min}$.
Note that $\hat{F}/\gamma$ decreases as $\hat{\phi}$ increases at fixed $\hat{R}$, as should be expected, but also as $\hat{R}$ increases at fixed $\hat{\phi}$; the latter occurs because an increase in obstacle size requires a correspondingly larger increase in cell size to keep $\hat{\phi}$ constant. Recall that the minimum attainable porosity is achieved for two distinct values of $\hat{R}$: $\hat{R}=\hat{R}_\text{crit}$ and $\hat{R}=1/2$, and with $\hat{a}=\hat{a}_\text{min}$ and although in that limit the porosity is equal, the maximum removal is nearly double in the case that $\hat{R}=\hat{R}_\text{crit}$ instead of the case where $\hat{R}=1/2$, highlighting the crucial role of microscale geometry in macroscopic transport and removal.

\subsection{Macroscale flow and transport properties}
\label{general_K_D_stuff}

For the specific geometry described above, we explore the impact of microstructure on macroscopic flow and dispersive transport by analysing the  permeability and effective  diffusivity tensors, $\hat{\bm{K}}$ (Figure~\ref{K_Dhat}) and $\hat{\bm{D}}$ (Figures~\ref{Pe=10_Dcomp}a,b and \ref{Pe=250_Dcomp}a,b)  and the physical components of $\hat{\bm{D}}$: $\hat{\bm{D}}_\text{obst}$ and $\hat{\bm{D}}_\text{disp}$ (Figures~\ref{Pe=10_Dcomp}c--f and \ref{Pe=250_Dcomp}c--f). Finally  we investigate the impact of continuously varying $\mathrm{Pe}_l$ on $\hat{\bm{D}}_\text{disp}$ (Figure~\ref{DdispVsPe}).
To determine $\hat{\bm{K}}$ and $\hat{\bm{D}}$ we solve Equations \eqref{stokes_homog_O_ep_scaled_naive} and 
 \eqref{Gamma_1}--\eqref{Gamma_cond}
  in COMSOL Multiphysics\textsuperscript{\textregistered}. We split the study into three solver steps because the flow problem can be solved independently first, before being used to solve the transport problem. These iterative steps generate a significant computational time saving.
 Firstly, we solve Equations \eqref{stokes_homog_O_ep_scaled_naive} 
   using Creeping Flow interface (`Fluid Flow' $\to$ `Single Phase Flow' $\to$ `Laminar flow (spf)') for both components of flow (\textit{cf.} \cite{auton2021homogenised}). Within the same code we feed this information into the  Coefficient Form PDE interface (`Mathematics'  $\to$ `PDE interfaces'  $\to$ `Coefficient form PDE')  and subsequently 
solve Equations  \eqref{Gamma_1}--\eqref{Gamma_cond} individually  for $\Gamma_1$ and subsequently $\Gamma_2$. 
This methodology  allows the partial differential equation (PDE) solvers to be provided with the converged creeping flow solutions initially. 
The domain is discretised using the `User-controlled mesh' $\to$ `Fluid dynamics', and with a boundary layer around all solid obstacles. The chosen mesh size  varies based on the miscroscale geometry.

 As in Auton \textit{et al}. \cite{auton2021homogenised}, we have two degrees of microstructural freedom, which allows us to explore the anisotropy in the system. However, unlike in Auton \textit{et al}. \cite{auton2021homogenised}, $\hat{\bm{D}}$ depends on Pe$_l\hat{\bm{\nabla}}\hat{P}$, where, without loss of generality, we fix $|\hat{\bm{\nabla}}\hat{P}|=1$. 
For arbitrary $\hat{\bm{\nabla}}\hat{P}$, with this geometry and any other geometry exhibiting at least two lines of symmetry through the cell centre that lie parallel to the $\hat{\bm{x}}$ axes, $\hat{\bm{K}}$ is diagonal but $\hat{\bm{D}}$ has non-zero off-diagonal components due to the dependence of $\hat{\bm{D}}_\text{disp}$ on Pe$_l\hat{\bm{\nabla}}\hat{P}$. In what follows, we restrict our investigations to the case when $\hat{\bm{\nabla}}\hat{P}= (-1,0)^\intercal$; in this limit both $\hat{\bm{K}}$ and $\hat{\bm{D}}$ are diagonal matrices.
 
\subsubsection{Comparison with limiting cases}
We validate our analysis for this geometry in several ways.  
First, in Figure~\ref{DdispVsPe} we demonstrate that the components of $\hat{\bm{D}}_\text{disp}$ each scale with $\mathrm{Pe}_l^2$, in agreement with the classic Taylor dispersion scaling  \cite{aris1956dispersion, davit2013homogenization, taylor1953dispersion}. Note also that $\left(\hat{\bm{D}}_\text{disp}\right)_{22}> \left(\hat{\bm{D}}_\text{disp}\right)_{11}$. This is because we use $\hat{a} = 1$ in these figures, corresponding to the hexagonal pattern of the obstacles being shorter in the $\hat{x}_1$ direction compared to the $\hat{x}_2$ direction. The aspect ratio of the hexagonal pattern is equal at $\hat{a} = \sqrt{3}$, and shorter in the $\hat{x}_2$ direction compared to the $\hat{x}_1$ direction for $\hat{a} > \sqrt{3}$.

\begin{figure}
    \centering
    \includegraphics[width=1\textwidth]{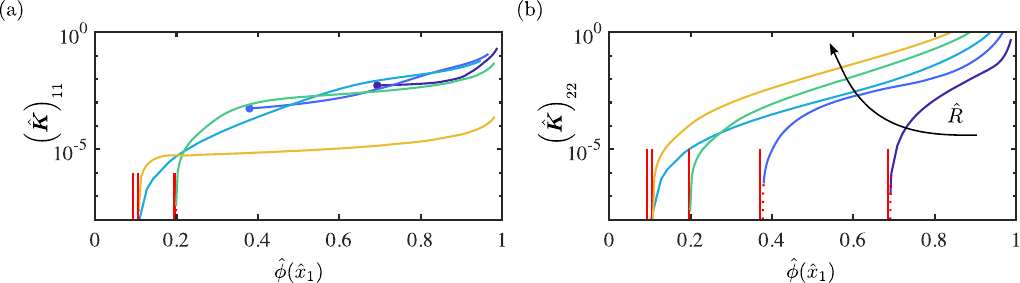}
    \caption{ \label{K_Dhat} The non-zero components of the effective permeability tensor against porosity for \mbox{$\hat{R}\in\{0.1, 0.2, \hat{R}_\text{crit}, 0.4, 0.49\}$} (blue to yellow; \textit{cf.} Eq.~\ref{Rcrit}), with $\hat{a}$ varying according to Equation~(\ref{phi_a_R}). 
 (a)~the effective longitudinal permeability $\left(\hat{\bm{K}}\right)_{11}$  and (b)~the effective transverse permeability $\left(\hat{\bm{K}}\right)_{22}$. The components of $\hat{\bm{K}}$ are independent of $\mathrm{Pe}_l$ and their difference implies macroscale anisotropy.   When $\hat{a}=\hat{a}_\text{min}(R)$ (Eq.~\ref{amin}), the behaviour of $\left(\hat{\bm{K}}\right)_{ii}$ depends on the value of $\hat{R}$: if $0<\hat{R}\leq \hat{R}_\text{crit}$ then  $\left(\hat{\bm{K}}\right)_{22}\to0$ whereas if  $\hat{R}<\hat{R}_\mathrm{crit}$ then both   $\left(\hat{\bm{K}}\right)_{11},\left(\hat{\bm{K}}\right)_{22}\to0$. The values to which $\left(\hat{\bm{K}}\right)_{ii}$ asymptote are shown as solid red vertical lines, and the interpolation between the last numerically obtained data point and the minimum obtainable porosity for the given $\hat{R}$ is shown as a dotted red line. 
    As $\hat{\phi} \to1$ ($\hat{a}\to\infty$) then  $\left(\hat{\bm{K}}\right)_{11}$ and $\left(\hat{\bm{K}}\right)_{22}$ diverge as the resistance to flow vanishes. 
  We fix $\hat{\bm{\nabla}}P= (-1,0)^\intercal$. }
\end{figure}

\begin{figure}
    \centering
        \includegraphics[width=1\textwidth]{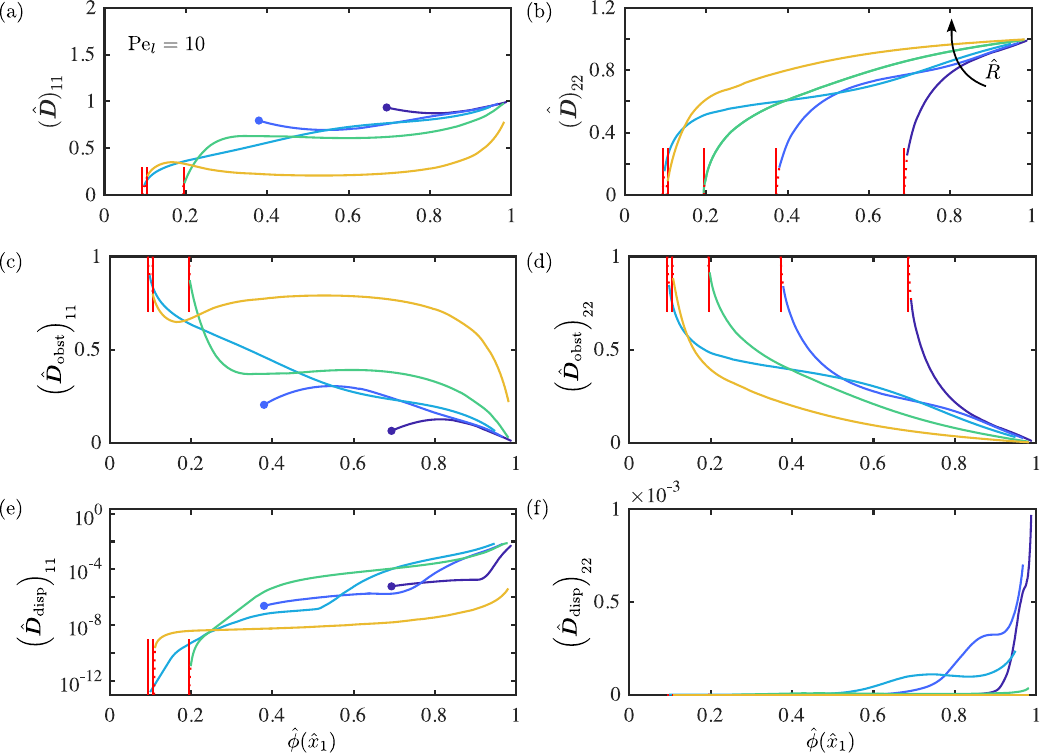}
    \caption{  \label{Pe=10_Dcomp} We investigate the behaviour of the non-zero components of the effective diffusivity $\hat{\bm{D}}$ and the contribution to  $\hat{\bm{D}}$ due to  the presence of obstacles $\hat{\bm{D}}_\mathrm{obst}$ and that resulting from dispersive effects $\hat{\bm{D}}_\mathrm{disp}$ (\textit{cf}. Eq.~\ref{Ds}) when $\mathrm{Pe}_l=10$. Note that, all quantities are plotted against $\hat{\phi}$, for the same fixed $\hat{R}$ values as in Figure~\ref{F}a and Figure~\ref{K_Dhat}. Further,  all the line colours and styles are the same as in Figure~\ref{F}a and Figure~\ref{K_Dhat}. In particular, note that the red vertical lines show the values to which the functions asymptote. 
   (a) the longitudinal effective diffusivity and (b) the  transverse effective diffusivity,  
    mirror the behaviour of $\left(\hat{\bm{K}}\right)_{ii}$ when $\hat{a}=\hat{a}_\text{min}$--- that is, when $\left(\hat{\bm{K}}\right)_{ii}$ vanishes, $\left(\hat{\bm{D}}\right)_{ii}$ also vanishes. 
     (c) the longitudinal component of  $\hat{\bm{D}}_\text{obst}$ and (d) the  transverse component of $\hat{\bm{D}}_\text{obst}$, represent the reduction of solute spreading due to the presence of obstacles; as expected when the presence of obstacles fully inhibits spreading both $\left(\hat{\bm{D}}_\text{obst}\right)_{11}$ and $\left(\hat{\bm{D}}_\text{obst}\right)_{22}$ tend to unity 
    which occurs when $\left(\hat{\bm{K}}\right)_{ii}\to0$. 
    (e) the longitudinal component of $\hat{\bm{D}}_\text{disp}$  and (f) the transverse component of $\hat{\bm{D}}_\text{disp}$, provide a measure for the enhancement of spreading due to shear forces;  $\left(\hat{\bm{D}}_\text{disp}\right)_{ii}$  vanishes when $\left(\hat{\bm{K}}\right)_{ii}\to0$.
    Recall that   
   $\hat{\bm{D}}_\mathrm{disp}$        is a function of       $\mathrm{Pe}_l\hat{\bm{\nabla}}\hat{P}$ and thus depends on the value of $\mathrm{Pe}_l$.  
   There is a 
   non-monotonicity in $\left(\hat{\bm{D}}_\mathrm{disp}\right)_{22}$ with respect to both $\hat{\phi}$ and $\hat{R}$ and $\left(\hat{\bm{D}}_\mathrm{disp}\right)_{11}>\left(\hat{\bm{D}}_\mathrm{disp}\right)_{22}$.
 }
\end{figure}

\begin{figure}
    \centering
        \includegraphics[width=1\textwidth]
        {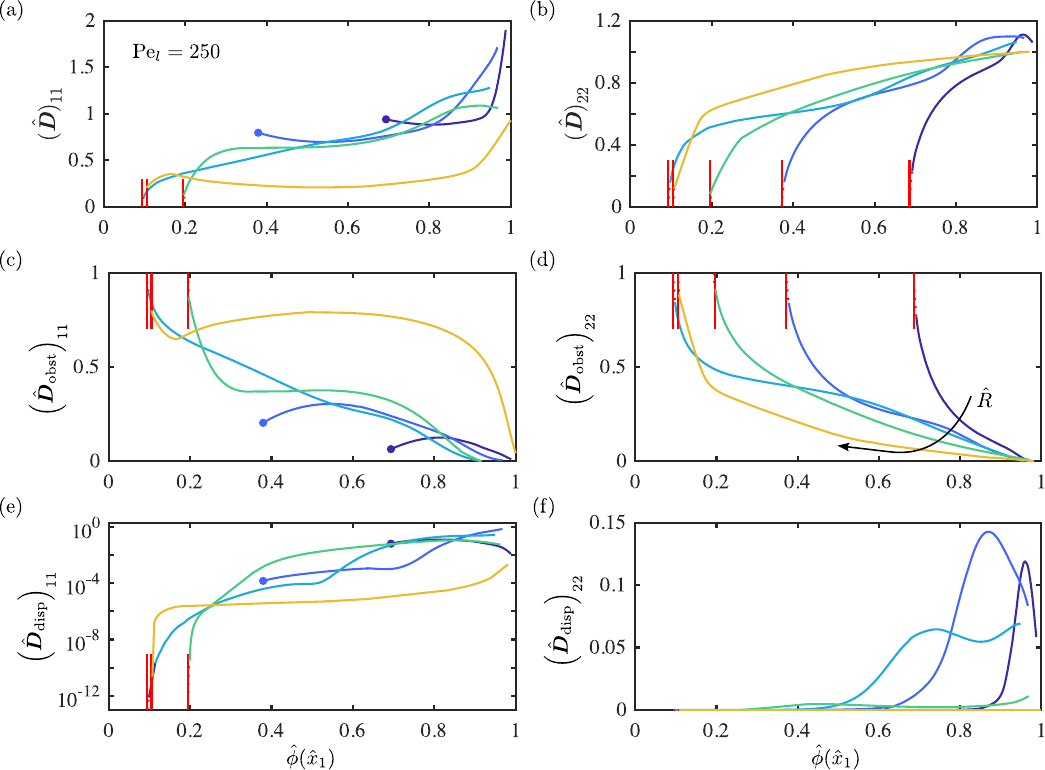}
    \caption{ \label{Pe=250_Dcomp} A replica of Figure~\ref{Pe=10_Dcomp}, but for $\text{Pe}_l=250$. We note the same limiting behaviour as Figure~\ref{Pe=10_Dcomp} when $\hat{a}=\hat{a}_\text{min}$, but note that the maximum magnitude of $\hat{\bm{D}}_\text{disp}$ is around two orders of magnitude greater  than the maximum magnitude of $\hat{\bm{D}}_\text{disp}$ with $\mathrm{Pe}_l=10$. 
    }
\end{figure}

Secondly, we have confirmed that the effective permeability and effective diffusivity vanish  when the filter loses connectivity ($\hat{a}\to\hat{a}_\text{min}(\hat{R})$). When $\hat{a}\to \hat{a}_\text{min}(\hat{R})$ with any fixed $\hat{R}<1/2$, the obstacles move closer together in the longitudinal direction  and the pore space becomes
disconnected in the transverse direction. Further,   for $\hat{R}\geq \hat{R}_\text{crit}$ the pore space is simultaneously  disconnected in the longitudinal direction.  Thus, as $\hat{a}\to \hat{a}_\text{min}(\hat{R})$, for all values of $\hat{R}$, $\left(\hat{\bm{K}}\right)_{22}$  and $\left(\hat{\bm{D}}\right)_{22}$ vanish and when $\hat{R}\geq \hat{R}_\text{crit}$, $\left(\hat{\bm{K}}\right)_{11}$  and $\left(\hat{\bm{D}}\right)_{11}$ vanish while for $\hat{R}< \hat{R}_\text{crit}$, $\left(\hat{\bm{K}}\right)_{11}$  and $\left(\hat{\bm{D}}\right)_{11}$ obtain their global minimum values in this limit but do not vanish as the longitudinal permeability does not vanish (Figures~\ref{K_Dhat}, \ref{Pe=10_Dcomp}a,b and \ref{Pe=250_Dcomp}a, b). When $R=1/2$, there is no transverse connectivity for all choices of $\hat{a}$. When connectivity is lost, the solid obstacles fully hinder the spreading of the solute so that  $\left(\hat{\bm{D}}_\text{obst}\right)_{ii}\to1$ and $\left(\hat{\bm{D}}_\text{disp}\right)_{ii}\to0$ (Figures~\ref{Pe=10_Dcomp}c--f and \ref{Pe=250_Dcomp}c--f), leading to an overall vanishing effective diffusivity.

Finally, we also see that the permeability diverges as the obstacles vanish ($\hat{\phi}\to 1$);  as $\hat{\phi} \to 1$ ($\hat{a}\to\infty$), both $\left(\hat{\bm{K}}\right)_{11}$ and $\left(\hat{\bm{K}}\right)_{22}$ diverge as the resistance to flow vanishes (Figure~\ref{K_Dhat}) and additionally $\left(\hat{\bm{D}}_\text{obst}\right)_{ii}\to0$ as molecular diffusion becomes unobstructed (Figures~\ref{Pe=10_Dcomp}c,d, \ref{Pe=250_Dcomp}c,d).

\subsubsection{Qualitative effect on solute transport of varying $\mathrm{Pe}_l$}

   \begin{figure}
    \centering
        \includegraphics[width=1\textwidth]{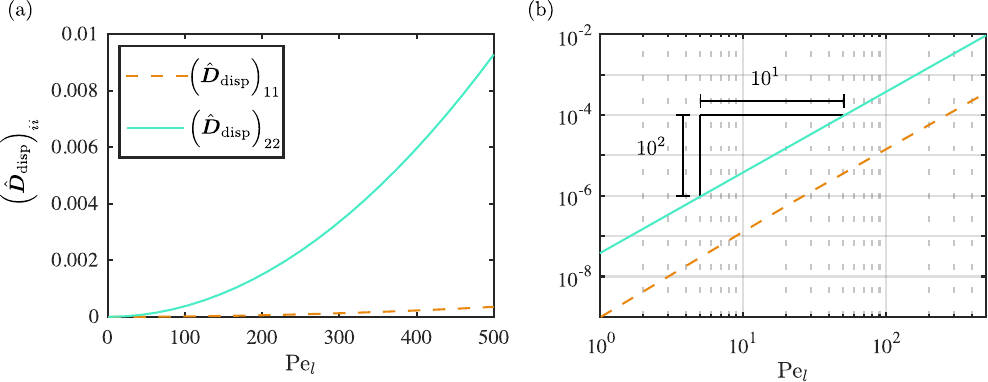}
    \caption{\label{DdispVsPe} The non-zero components of 
    the effective diffusivity: the longitudinal diffusivity $\left({\hat{\bm{D}}_\text{disp}}\right)_{11}$ (dashed line, orange) and  the transverse diffusivity $\left(\hat{\bm{D}}_\text{disp}\right)_{22}$ (solid line, aqua) against $\mathrm{Pe}_l$, when $\hat{a}\equiv1$ and $\hat{\bm{\nabla}}\hat{P}= (-1,0)^\intercal$. (a) Both $\left(\hat{\bm{D}}_\text{disp}\right)_{11}$ and $\left(\hat{\bm{D}}_\text{disp}\right)_{22}$ increase monotonically with $\mathrm{Pe}_l$. (b) $\left(\hat{\bm{D}}_\text{disp}\right)_{ii}$ is shown on a log--log scale, which shows  that $\left(\hat{\bm{D}}_\text{disp}\right)_{22}>\left(\hat{\bm{D}}_\text{disp}\right)_{11}$ and $\left(\hat{\bm{D}}_\text{disp}\right)_{ii}\propto$~Pe$_l^2$~\cite{aris1956dispersion, davit2013homogenization, taylor1953dispersion}. 
    }
\end{figure}

For the majority of the $\hat{a}$--$\hat{R}$--$\hat{\phi}$ parameter space, the magnitude of $\hat{\bm{D}}$ is greater when $\mathrm{Pe}_l=250$ (Figure~\ref{Pe=250_Dcomp}) than when $\mathrm{Pe}_l=10$ (Figure~\ref{Pe=10_Dcomp}), 
This is largely explained by the $\text{Pe}_l^2 $ scaling for large $\text{Pe}_l$ discussed above, which increases the relative importance of the dispersive component of the effective diffusivity (Figure~\ref{DdispVsPe}). 

However, the qualitative behaviour of $\hat{D}$  is similar for both values of $\text{Pe}_l$.
The main qualitative difference  in  $\left(\hat{\bm{D}}\right)_{11}$  between the two values of $\mathrm{Pe}_l$ is in the behaviour of the $\hat{R}=0.4$ curve;  when $\text{Pe}_l=10$ as $\hat{\phi}\to1$, we have an apparent monotonic increase of $\left(\hat{\bm{D}}\right)_{11}$, while for $\text{Pe}_l=250$, there is a local maxima for $\hat{\phi}\sim0.9$ ($\hat{a}\sim10)$.   For $\hat{R}\geq \hat{R}_\text{crit}$,   $\left(\hat{\bm{D}}\right)_{22}$ increases monotonically from its minimum value  to its maximum value as $\hat{\phi}\to1$ for both values of $\mathrm{Pe}_l$. However, for  $\hat{R}< \hat{R}_\text{crit}$, when $\mathrm{Pe}_l=10$,  $\left(\hat{\bm{D}}\right)_{22}$ also undergoes the same monotonic behaviour, while when $\mathrm{Pe}_l=250$, $\left(\hat{\bm{D}}\right)_{22}$ obtains its maximum value for some $\hat{\phi}<1$.  
 This non-monotonicity suggests that there is some microscale geometry that optimises the dispersive transport.  

Qualitatively both $\left(\hat{\bm{D}}_\text{obst}\right)_{11}$ and $\left(\hat{\bm{D}}_\text{obst}\right)_{22}$ are very similar for the cases where $\mathrm{Pe}_l=10$ and $\mathrm{Pe}_l=250$  for all $\hat{R}$. 
This is because the impingement on  spreading due to the presence of obstacles becomes negligible as the obstacles become arbitrarily spaced in the longitudinal direction.
The only substantive difference in $\left(\hat{\bm{D}}_\text{obst}\right)_{11}$ between the $\mathrm{Pe}_l=10$ and $\mathrm{Pe}_l=250$ cases is in the non-monotonic behaviour as $\hat{R}$ varies for large fixed $\hat{\phi}$. For example, when $\hat{\phi}=0.8$, in the case that $\mathrm{Pe}_l=10$ we have that $\left(\hat{\bm{D}}_\text{obst}\right)_{11}$ increases monotonically with increasing $\hat{R}$ whereas   for $\mathrm{Pe}_l=250$ the  smallest  value of $\left(\hat{\bm{D}}_\text{obst}\right)_{11}$ occurs for $\hat{R}=\hat{R}_\text{crit}$. 

 We expect to see non-monotonicities  when the dispersive component dominates, which corresponds to high $\text{Pe}_l$; this is because of the competing transport mechanisms \cite{liu2024non}. At small $\mathrm{Pe}_l$, diffusion dominates, while as $\mathrm{Pe}_l$  increases the different dispersive mechanisms become increasingly important. 
In particular, as well as the classical   dispersion caused by non-uniform velocity profiles within pores or throats, there are also effects due to hold-up dispersion. This occurs due to areas of low flow and mechanical dispersion  resulting from the repeated separation and merging of flow passages at the junctions of the pore space \cite{liu2024non}. For $\mathrm{Pe}_l=10$, the contribution  to $\hat{\bm{D}}$ from dispersion is negligible (Figure~\ref{Pe=10_Dcomp}e,f) and thus  as $\hat{\phi}\to 1$, $\left(\hat{\bm{D}}\right)_{ii}\sim1$ (Figure~\ref{Pe=10_Dcomp}a,b).
However, for $\mathrm{Pe}_l=250$ the dispersive effects in the limit $\hat{\phi}\to 1$ are $\mathcal{O}(1)$ (Figure~\ref{Pe=250_Dcomp}a,b) and thus $\left(\hat{\bm{D}}\right)_{ii}$ is increased beyond unity (Figure~\ref{Pe=250_Dcomp}e,f). In particular,  Figure~\ref{Pe=250_Dcomp}b shows a non-monotonicity in $(\hat{\bm{D}})_{22}$  --- that is, the maximum transverse spreading is not achieved in the limit $\hat{\phi}\to1$, but rather for some combination of $\hat{a}$ and  $\hat{\phi}<1$. 
Figure~\ref{Pe=250_Dcomp}f, highlights the origin of  this non-monotonicity; we see that it is  solely due to $\left(\hat{\bm{D}}_\text{disp}\right)_{22}$, as $\left(\hat{\bm{D}}_\text{obst}\right)_{22}$ shows a clear monotonic decrease as $\hat{\phi}$ increases. 
Further, although no non-monotonic behaviour is seen in $\left(\hat{\bm{D}}\right)_{22}$ when $\mathrm{Pe}_l=10$ (Figure~\ref{Pe=10_Dcomp}b), in this case $\left(\hat{\bm{D}}_\text{disp}\right)_{22}$  shows a non-monotonicity with increasing $\hat{\phi}$ (Figure~\ref{Pe=10_Dcomp}f). 

\section{Conclusions}
\label{s:conclusion}

We have presented a formal derivation for dispersive transport within a heterogeneous porous medium comprising cells of varying size
each containing multiple arbitrarily shaped obstacles, for a general, incompressible,  fluid flow \cite{allaire2010two, allaire2007homogenization}. 
We considered an advection-dominating regime, which introduces dispersion into the problem; this is important in many industrial filtration scenarios, where dispersion becomes important \cite{bedrikovetsky2006correction}. 
The homogenisation was conducted using both multiple spatial and temporal scales, enabling us to deal with both the dispersive limit \cite{salles1993taylor} and the microscale heterogeneity \cite{auton2021homogenised}. 
The heterogeneity within the porous material originates from slowly varying obstacle size and/or obstacle spacing along the length of the porous 
medium; the latter also induces strong anisotropy within the problem. This results in a near-periodic microscale problem; importantly, the variation in spacing means that the period of the microscale depends on the macroscale, which means that the upscaling of this problem requires a nontrivial modification of classic homogenisation. This builds on our previous work for slower flows with no dispersive effects~\cite{auton2021homogenised}.
To account for the dominant advective terms during the homogenisation, we introduced a second, fast temporal scale following the methodology presented in Salles \textit{et al.}~\cite{salles1993taylor}.
The fast timescale highlights how the transport equation is dominated by advection at leading order --- that is, the solute advects (convects) with the fast fluid flow. We use this result to eliminate terms at higher orders and, on recombining the two temporal scales, we determine the leading-order transport equation on the slow timescale.
This  homogenised equation is an advection--diffusion--reaction equation, in which advection dominates, with an anisotropic effective diffusivity tensor, which itself comprises components due to molecular diffusivity, a reduction in spreading due to the presence of obstacles ($\hat{\bm{D}}_\text{obst}$), and a dispersive component  ($\hat{\bm{D}}_\text{disp}$), which depends on the product of the local P\'{e}clet number, $\mathrm{Pe}_l$, and pressure gradient, $\hat{\bm{\nabla}}\hat{P}$. 
The permeability, effective diffusivity and the removal terms are functions of the porosity, obstacle spacing and a scale factor controlling the variation in obstacle size across the medium; any two of these are free choices, which prescribe the third. 
To deal with the dominance of advection, we performed a subsequent drift transform 
to the frame of reference  moving with the solute pulse. Under this transformation, a careful application of Taylor's expansions provides the leading-order equation that governs the spreading of the solute (\textit{cf}.~Eq.~\ref{MMS_then_drift}). The resulting macroscale equations are computationally inexpensive to solve, allowing for optimisation of parameters through large sweeps, which would not be possible with direct numerical simulation (DNS). 

In \S\ref{exs} we considered a simple geometry comprising circular obstacles in a hexagonal array and fixed a particular incompressible fluid flow: Stokes flow.  
We determined the corresponding permeability and effective diffusivity numerically, and show how these depend on the radius of the obstacles and the aspect ratio of the cells. 
This work illustrates and quantifies how the permeability, diffusivity  and dispersivity of a porous medium depend not only on the porosity of the medium, but also on its microstructure and the magnitude and direction of the driving pressure gradient. 
The dispersive component of the effective diffusivity is shown to be proportional to $\mathrm{Pe}_l^2$ (Figure~\ref{DdispVsPe}), in agreement with the classical Taylor dispersion scaling~\cite{aris1956dispersion, taylor1953dispersion}. 

While we focus on the two dimensional case here for clarity in dealing with the non-standard homogenisation approach, we note that it is straightforward to generalise our results to the three-dimensional problem. We would expect the three-dimensional results to be qualitatively similar to the two-dimensional problem considered here in general, with the important exception of a non-vanishing connectivity when obstacles touch.
As in Auton \textit{et al}. \cite{auton2021homogenised},  we have assumed that the solute particles are negligibly small. If one were explicitly interested in understanding the effect of the smallest distances between adjacent obstacles (choke points), finite-size effects of particles may need to be considered, including the subsequent effect on geometry.  The freedom in the microscale geometry allows for the construction of a porous structure with sufficiently wide longitudinal connectivity as to avoid blockages. Further, we note that applying a stress to the porous medium may vary the spacing between obstacles. This could be accounted for by coupling our model to an appropriate stress-strain relationship. 

We have validated our results against limiting cases; DNS for flow and transport in a broader range of relevant geometries would provide further validation and may lead to additional insight, and should be the subject of future work. 
Here, we have limited our consideration to the case when the pressure gradient is purely longitudinal (\textit{i.e.,} $\hat{\bm{\nabla}}\hat{P}\equiv(-1,0)^\intercal$), however further investigation into the cases where 
$\hat{\bm{\nabla}}\hat{P}\neq(-1,0)^\intercal$ is warranted and will undoubtedly yield further insights into the effect of microscale heterogeneity on macroscopic dispersive transport and removal. We also note that, the details of the behaviour of $\hat{\bm{D}}_\text{disp}$ in the limit $\hat{a}\to\infty$ require more careful investigation as $\mathrm{Pe}_l$ increases.

In summary, the results presented in this manuscript form a comprehensive framework for describing the macroscropic dispersive transport and removal properties of a heterogeneous porous medium, subject to a general, incompresssible flow. 
\section{Acknowledgments}

LCA  acknowledges the CERCA Programme
of the Generalitat de Catalunya for institutional support, their work was also supported by the Spanish State Research Agency,
through the Severo Ochoa and Maria de Maeztu Program for Centres and Units of Excellence in R\&D (CEX2020-001084-M)  and has been partially funded by MCIN/AEI/ 10.13039/50110 0 011033/ and by `ERDF A way of making Europe', grant number PID2020- 115023RB-I00. 

\appendix

\section{Uniqueness Proof} 
\label{Uniqueness}

We consider $\bm{y}\in\omega_f$, the transformed microscale $\bm{y}$-cell and we define $\partial\omega_f\defeq \partial\omega \cup\left(-\partial\omega_s\right)$ to be the boundary to the fluid region. Over this domain Equations~(\ref{ad_diff_leadingorder_Mo}) 
 of the manuscript
are valid: 
\begin{subequations}\label{ad_diff_leadingorder_MoAp}
 \begin{equation}
 \label{app1}
\bm{0}=  \bm{\nabla}_y^a\cdot\left(\bm{\nabla}_y^a c^{(0)}-\mathrm{Pe}_l\ \bm{v}^{(0)}c^{(0)}\right), \ \quad {\bm{y}}\in{\omega}_f(x_1),
        \end{equation}
       \begin{equation}
       \label{app2}
        \bm{0}= \left(\bm{\sigma}\cdot\bm{n}^y\right)
        \cdot
        \left(\bm{\nabla}_y^a c^{(0)} -\mathrm{Pe}_l\ \bm{v}^{(0)}c^{(0)}\right)
        , \quad {\bm{y}}\in \partial{\omega}_s(x_1), 
\end{equation}
and 
\begin{equation}
\label{app3}
\bm{v}^{(0)}, \quad c^{(0)}, \quad \text{periodic on} \quad \bm{y}\in\partial\omega_{=}
(x_1)\ \ \text{and}\ \  \partial\omega_{||}(x_1),
\end{equation}
\end{subequations}
with the tensor $\bm{\sigma}$ defined in Equation~(\ref{sigma}). 
Recall that $\bm{v}^{(0)}$ is a known function of $\bm{x}$ and $\bm{y}$, subject to $\bm{\nabla_y^a}\cdot\bm{v}^{(0)}=0$ over $\omega_f$, while $c^{(0)}$ is an unknown function of $\bm{x}$, $\bm{y}$, $t$ and $\tau$. Thus, Equation~(\ref{app1}) can be expressed as 
\begin{equation}
    \left(\nabla_y^a\right)^2c^{(0)} - \mathrm{Pe}_l\bm{v}^{(0)}\cdot\bm{\nabla_y^a}c^{(0)} = 0, \ \quad {\bm{y}}\in{\omega}_f(x_1),
\end{equation}
from which we can clearly see that $c^{(0)}(\bm{x},\bm{y},t,\tau) = c^{(0)}(\bm{x},t,\tau)$ is \textit{a} solution. Our aim remains to show that this is \textit{the only} solution subject to Equations~(\ref{app2})--(\ref{app3}). Recalling that there is a non-slip and no-penetration boundary condition on the fluid velocity at all orders (Eq.~\ref{vi}) 
, Equations~(\ref{app2}) can be expressed as 
\begin{equation}
\bm{0}=\left(\bm{\sigma}\cdot\bm{n}^y\right)\cdot\bm{\nabla}_y^ac^{(0)},  \ \quad {\bm{y}}\in{\partial\omega}_s(x_1).
\end{equation} Thus our problem becomes 
\begin{subequations}
\begin{equation}
\label{soln_1}
 \left(\nabla_y^a\right)^2c^{(0)}- \bm{\nabla}_y^a\cdot(\bm{v}^{(0)}c^{(0)}) = 0 \quad \text{and } \quad \nabla\cdot\bm{v}^{(0)}=0 \quad \text{for} \quad \bm{x}\in \omega
 \end{equation}
subject to 
\begin{equation}
\label{sigmany}
\left(\bm{\sigma}\cdot\bm{n}^y\right)\cdot\bm{\nabla}_y^ac^{(0)}=\bm{0} \quad \text{and} \quad \bm{v}^{(0)}=0 \quad \text{for}\quad \bm{y}\in\partial\omega_s
\end{equation}
\end{subequations}
and periodicity of $\bm{v}^{(0)}$, $c^{(0)}$ and $\bm{\nabla}_y^a c^{(0)}$ on $y\in\partial\omega_{||}$ and $\partial\omega_{=}$,  respectively. We also assume that $c^{(0)}\in C^1$ and that $c^{(0)}$ is bounded.

Multiplying Equation~(\ref{soln_1}) by $c^{(0)}$ and integrating over $\omega\defeq\omega_f\cup\omega_s$ yields 
\begin{multline}
\label{int_1}
 0 = -\int_\omega c^{(0)}\left(\nabla_y^a\right)^2c^{(0)}\  \mathrm{d}S_y+\int_\omega c^{(0)}\bm{\nabla}_y^a\cdot (\bm{v}^{(0)}c^{(0)})\ \mathrm{d}S_y  \\ =\int_\omega|\bm{\nabla}_y^a c^{(0)}|^2\ \mathrm{d}S_y-\underbrace{
 \int_{\partial\omega_f}c^{(0)}\bm{\nabla}_y^a c^{(0)}\cdot\bm{n}^y_f\ \mathrm{d}s_y
 }_{(\star)} + \underbrace{
  \int_\omega c^{(0)}\bm{\nabla}_y^a\cdot(\bm{v}^{(0)}c^{(0)})\ \mathrm{d}S_y
 }_{(\square)},
\end{multline}
where $\bm{n}_f^y$ is the unit outward-facing normal to the fluid region and where equality comes from use of the Divergence Theorem and from the identity
\begin{equation}
c^{(0)}\left(\nabla_y^a\right)^2c^{(0)} = \bm{\nabla}_y^a\cdot(c^{(0)}\bm{\nabla}_y^a c^{(0)})-|\bm{\nabla}_y^a c^{(0)}|^2,
\end{equation}
and where $\bm{n}^y_f\defeq\bm{n}^\square\cup\left(-\bm{n}_s\right)$.

 We now show that the last two terms on the right-hand side of  Equation~(\ref{int_1}),   $(\star)$ and $(\square)$, are both individually equal to 0;	
we separately consider $\partial\omega_s$, $\partial\omega\defeq\partial\omega_{||} \cup\partial\omega_{=}$. Recall  from Equation~(3.7a) of the manuscript
that 
\begin{equation}
\label{n_s}
    \bm{n}_s\sim \frac{\sigma_{ij}n_j^y\bm{e}_i}{(\sigma_{kl}\sigma_{km}n_l^yn_m^y)^{1/2}} \equiv \frac{\bm{\sigma}\cdot\bm{n}^y}{(\sigma_{kl}\sigma_{km}n_l^yn_m^y)^{1/2}}.
\end{equation}
We decompose $(\star)$:
\begin{equation}
     \int_{\partial\omega_f}c^{(0)}\bm{\nabla}_y^a c^{(0)}\cdot\bm{n}^y_f\ \mathrm{d}s_y =      \int_{\partial\omega}c^{(0)}\bm{\nabla}_y^a c^{(0)}\cdot\bm{n}^\square\ \mathrm{d}s_y-\int_{\partial\omega_s}c^{(0)}\bm{\nabla}_y^a c^{(0)}\cdot\bm{n}_s\ \mathrm{d}s_y.
\end{equation}
The first term on the right-hand side vanishes due to the periodicity of $c^{(0)}$ and \linebreak Equation~(\ref{n_s}) gives 
\begin{equation}
\label{star2}
\int_{\partial\omega_s}c^{(0)}\bm{\nabla}_y^a c^{(0)}\cdot\bm{n}_s\ \mathrm{d}s_y\sim     \int_{\partial\omega_s} \left(\frac{c^{(0)}}{(\sigma_{kl}\sigma_{km}n_l^yn_m^y)^{1/2}}\right)(\bm{\sigma}\cdot\bm{n}^y)\cdot\bm{\nabla}_y^a c^{(0)} \ \mathrm{d}s_y.
\end{equation}
Thus by the solid boundary condition~(Eq.~\ref{sigmany}), the right-hand side of Equation~(\ref{star2}) vanishes which shows that, at leading order,  $(\star)$ vanishes. 

Consider the identities
\begin{subequations}
\label{identities}
\begin{equation}
c^{(0)}\bm{\nabla}_y^a\cdot(\bm{v}^{(0)}c^{(0)}) = (\bm{v}^{(0)}\cdot\bm{\nabla}_y^a c^{(0)})c^{(0)}+(\bm{\nabla}_y^a\cdot\bm{v}^{(0)})(c^{(0)})^2 = (\bm{v}^{(0)}\cdot\bm{\nabla}_y^a c^{(0)})c^{(0)}
\end{equation}
where we have recalled that microscale incompressibility (Eq.~\ref{n transform in geo})
, and similarly,
\begin{equation}
c^{(0)}\bm{\nabla}_y^a\cdot(\bm{v}^{(0)}c^{(0)}) = \bm{\nabla}_y^a\cdot(\bm{v}^{(0)}(c^{(0)})^2)- (\bm{v}^{(0)}\cdot\bm{\nabla}_y^a c^{(0)})c^{(0)}, 
\end{equation}
\end{subequations}
Thus, combining Equations~(\ref{identities})  yields 
$c^{(0)}\bm{\nabla}_y^a\cdot(\bm{v}^{(0)}c^{(0)}) = \frac{1}{2}\bm{\nabla}_y^a\cdot(\bm{v}^{(0)} (c^{(0)})^2)$. 
Hence, on  application of the Divergence Theorem to $(\square)$ gives 
\begin{equation}
 \int_\omega c^{(0)}\bm{\nabla}_y^a\cdot(\bm{v}^{(0)}c^{(0)})\ \mathrm{d}S_y = \frac{1}{2}\int_{\partial\omega_f}(\bm{v}^{(0)}(c^{(0)})^2)\cdot\bm{n}_f^y\ \mathrm{d}s_y.
\end{equation}
Once again we split this into two cases:  on $\partial\omega_s$ we have that $\bm{v}^{(0)}=0$ and $c^{(0)}$ is bounded and thus 
\begin{subequations}
\begin{equation}
    \int_{\partial\omega_s}(\bm{v}^{(0)}c^2)\cdot\bm{n}_s\ \mathrm{d}s_y = 0; 
\end{equation}
while on $\partial\omega_{||}$ and  $\partial\omega_{=}$ 
\begin{equation}
    \int_{\partial\omega}(\bm{v}^{(0)}c^2)\cdot\bm{n}^\square\ \mathrm{d}s_y = 0;
\end{equation}
\end{subequations}
since both $c^{(0)}$ and $\bm{v}^{(0)}$ are  periodic.  Thus $(\square)$ vanishes also. 
Hence, Equation~(\ref{int_1}) becomes
\begin{equation}
0= \int_\omega|\bm{\nabla}_y^a c^{(0)}|^2\ \mathrm{d}S_y
 \end{equation}
 which tells us that $c^{(0)}(\bm{x},\bm{y},t, \tau)\equiv C^{(0)}(\bm{x},t, \tau)$ such that $c^{(0)}= C^{(0)}$ is independent of the microscale for all $\bm{y}\in\omega$.

\bibliographystyle{plain}
\bibliography{references}
\end{document}